\documentclass[a4paper,showpacs,preprintnumbers,nofootinbib,superscriptaddress,groupedaddress,11pt]{article}
\usepackage[utf8]{inputenc}
\usepackage{cite}
\usepackage[english]{babel}
\usepackage[nottoc]{tocbibind}
\makeatletter
\def\l@subsubsection#1#2{}
\def\l@subsubsubsection#1#2{}
\makeatother

\usepackage{subfigure}
 \usepackage{yfonts}
\usepackage{graphicx,amssymb,amsmath,amsthm,amsfonts,amscd, mathrsfs,epsfig,epsf}
\usepackage[mathcal]{euscript}
\usepackage[usenames]{color}
\usepackage{epstopdf}
\definecolor{darkred}{rgb}{0.5,0,0}
\usepackage{latexsym}
\usepackage{array}
\usepackage{afterpage}
\usepackage{bm}
\usepackage{dcolumn}
\usepackage[utf8]{inputenc}
\usepackage{rotating}
\usepackage{longtable}

\usepackage{enumerate}
\usepackage{tensor,multirow}
\usepackage{url}
\usepackage{placeins}
\usepackage[linktocpage]{hyperref}
\usepackage{float}
\usepackage{mwe}
\def\be{\begin{eqnarray}}
\def\ee{\end{eqnarray}}
\usepackage{soul}

\newcounter{mnotecount}[section]

\renewcommand{\themnotecount}{\thesection.\arabic{mnotecount}}

\newcommand{\mnote}[1]
{\protect{\stepcounter{mnotecount}}$^{\mbox{\footnotesize
$
\bullet$\themnotecount}}$ \marginpar{
\raggedright\tiny\em
$\!\!\!\!\!\!\,\bullet$\themnotecount: #1} }

\title{Quasinormal modes and stability of boosted Reissner-Nordstr\"om AdS black holes}

\author{Rodrigo D. B. Fontana$^{(1)}$ and Filipe C. Mena$^{(2,3)}$ \\\\
{\small $^{(1)}$Universidade Federal do Rio Grande do Sul, Campus Tramanda\'i-RS}
\\
{\small Estrada Tramanda\'{i}-Os\'orio, CEP 95590-000, RS, Brazil}
\\
{\small $^{(2)}$Centro de An\'alise Matem\'atica, Geometria e Sistemas Din\^amicos,}
\\
{\small Instituto Superior T\'ecnico, Univ. Lisboa, Av. Rovisco Pais 1, 1049-001 Lisboa, Portugal}
\\
{\small $^{(3)}$Centro de Matem\'atica, Universidade do Minho, 4710-057 Braga, Portugal}
}

\date{\today}
\begin{document}
\maketitle
\begin{abstract}

We investigate the numerical stability of accelerated AdS black holes against linear scalar perturbations. In particular, we study the evolution of a probe non-minimally coupled scalar field on Schwarzschild and Reissner-Nordström AdS black holes with small accelerations by computing the quasinormal modes of the perturbation spectrum. 
We decompose the scalar field Klein-Gordon equation and study the eigenvalue problem for its angular and radial-temporal parts using different numerical methods. 
 The angular part is written in terms of the Heun solution and expanded through the Frobenius method which turns out to give eigenvalues qualitatively similar to the ones obtained through the spherical harmonics representation. The radial-temporal evolution renders a stable field profile which is decomposed in terms of damped and purely imaginary oscillations of the quasinormal modes. We calculate the respective frequencies for different spacetime parameters showing the existence of a fine-structure in the modes, for both real and imaginary parts, which is not present in the non-accelerated AdS black holes. Our results indicate that the Schwarzschild and Reissner-Nordström AdS black holes with small accelerations are stable against linear scalar perturbations.

\end{abstract}

\section{Introduction} 

Boost symmetric geometries of General Relativity may provide models for accelerated black holes and have been an active area of research, in particular in studies about the creation of black hole pairs \cite{Hawking-Ross, Hawking-Horowitz-Ross}, the splitting of cosmic strings \cite{Eardley} and, notably, the discovery of the black ring solution in five dimensions \cite{Emparan-Reall}. Moreover, interesting recent works provide a solid ground for the thermodynamics of accelerated black holes \cite{Appels_2016} as well as developments about their observational aspects such as the existence of shadows \cite{zhang2020shadows} and gravitational lensing \cite{frost2020lightlike}. 

The stability of accelerated black holes is then an important question to investigate. However most of the studies in this direction have so far been restricted to spacetimes without a cosmological constant (see \cite{destounis2020stability} and references therein). 
In this paper, motivated by the important role of the AdS-CFT correspondence in physics, we investigate the numerical stability of boosted black hole solutions with a negative cosmological constant. In particular, we consider a non-minimally coupled scalar field perturbation on boosted AdS black hole backgrounds and study the evolution of quasinormal modes associated to the Klein-Gordon field equation.

The boosted AdS black hole spacetimes have been investigated in  \cite{Plebanski:1976gy, Podolsky:2002nk, Dias_2003} where it was realized that the conformal properties of the metric may drastically vary according to the boost parameter. For small boosts, the asymptotic conditions from the point of view of the wave propagation problem are similar to the non-accelerated AdS black holes. On the other hand, for large accelerations, the presence of an accelerated horizon demands different boundary conditions \cite{Destounis2020pjk}.

Considering the existence of an acceleration parameter in the metric, the causal geodesics' paths are significantly altered \cite{lim2020null} as well as the shadow radius around the black holes \cite{zhang2020shadows} and their gravitational lensing physics \cite{frost2020lightlike} thus providing an experimental arena for observational tests.
The black hole thermodynamics is also non-trivially affected by the presence of the conical singularities \cite{Appels_2016,Gregory_2017}. Such structures generate a topological tension in the spacetime, proportional to the acceleration parameter that contributes to the energy of the black hole, via Smarr's formula. This affects, as well, other canonical thermodynamical quantities like the entropy, temperature and surface gravity at the horizons. So, an interesting question is how such non-trivial differences change the response of the geometry to probe fields. 

Relative to the AdS-CFT correspondence, the linear perturbations of black holes with a negative cosmological constant on the gravitational side have an interesting interpretation in terms of the conformal field theory. For large black holes, the imaginary part of the quasinormal mode represents the inverse of the timescale for equilibrium in the associated field theory \cite{Horowitz_2000} and the temperature scales directly the imaginary part of the frequencies. In turn, for small black holes, the damping of oscillation is proportional to the event horizon and, for BTZ-like black holes, the poles of the correlation function living in a (1+1)-dimensional field theory are exactly those field oscillation modes of the black hole. This provided a new interpretation of those frequencies in terms of its duals \cite{Birmingham_2002}. 

From the mathematical point of view, the stability of the non-accelerated Schwarzschild AdS spacetime for the spherically symmetric Einstein-Klein-Gordon system has been proved by \cite{holzegel2, holzegel} (see also \cite{holzegel3,holzegel4} for stability results in the case of Kerr AdS). More recently, \cite{kehle} proved results about the stability of solutions to the Klein-Gordon equation in Reissner-Nordstr\"om AdS black holes. Interestingly, \cite{kehle2} showed that for Reissner-Nordstr\"om AdS black holes, the $C^0$-formulation of the linear analog of the strong censorship conjecture is false but the $H^1$-formulation is true. It is conceivable that the mathematical results about stability can be extended to the accelerated versions of the AdS black holes but that remains to be seen. 

Regarding numerical results, there have been works exploring the numerical stability of non-accelerated Schwarzschild AdS \cite{Horowitz_2000, Cardoso_2003, Konoplya_2002} as well as Reissner-Nordstr\"om AdS black holes by computing the quasinormal modes of scalar field perturbations \cite{Wang:2004bv, Wang:2000dt}.

Our work can be seen as generalisation of the latter numerical studies by including an acceleration parameter. In particular, we investigate the stability of accelerated AdS black holes by studying the evolution of a probe scalar field propagating on AdS black hole backgrounds with and without charge. 

The dynamical response of the geometry to the field is expected to be that of a tower (or multiple towers) of quasinormal modes. Those frequencies represent transient solutions from the scattered wave of the scalar field potential. By means of the potential analysis, we will investigate possible unstable or stable spacetimes, the latter having always positive potentials. In the case of (partially) negative potentials, instabilities may occur and further analysis is needed. In that case, the study of the characteristic integration may provide a strong indication about the stability of the geometry when the field profile decays in time. 
Once the scalar field signal is acquired, the quasinormal modes will be extracted with spectral techniques applied to the field profile. 

The paper is organized as follows, in sections \ref{s2} and \ref{s3} we revise the boosted AdS black holes with charge and acceleration. In section \ref{s4} we summarize the numerical methods and the field ansatz used for the reduction of the Klein-Gordon equation into a separable system. In section \ref{s5} we obtain the frequencies resulting from the numerical integration identifying some of the interesting features of the scalar field perturbation such as the existence of a fine structure with varying azimuthal momentum, which is not present in non-accelerated cases, as well as the presence of purely imaginary modes. We present our conclusions in section \ref{s6}.

\section{Boosted Reissner-Nordström AdS metrics}\label{s2}

In this section we present the metric forms that we will use and the coordinate transformations between them. This will allow us to keep track of the relevant physical parameters involved in the metrics and clarify the geometric properties of the solutions. 

We consider boosted black hole solutions given by the $C$-metric with charge, mass and a negative cosmological constant, henceforth referred as boosted Reissner-Nordström Anti-de Sitter (BRNAdS) spacetimes (see e. g. \cite{Podolsky:2002nk, Krtou__2005, Dias_2003, Podolsk__2003, Podolsk__2001}): 
\be
ds^2 = W^{-2} \left( - Fdt^2 + \frac{1}{F}dy^2 + \frac{1}{G}dx^2 + G d\Phi^2 \right),
\label{e1}
\ee
with
\begin{equation}
\begin{aligned}
\label{e2}
W & =  a(x+y) \\
F & =   \frac{1}{a^2 l^2} - 1 + y^2 - 2a m y^3 + a^2q^2y^4\\
G & =  1 - x^2 - 2a m x^3 - a^2q^2x^4,
\end{aligned}
\end{equation}
where $a, m, q$  and $l$ are constant parameters. As an intermediate step we note that such metric can also be written in the form (see e. g. \cite{Griffiths:2006tk,Hong_2003,Kofro__2015,Bi_k_2009,Kofro__2016})
\be
ds^2 = \textgoth{W}^{-2} \left( - \textgoth{F}d\tau^2 + \frac{1}{\textgoth{F}}dY^2 + \frac{1}{\textgoth{G}}dX^2 + \textgoth{G} d\phi^2 \right),
\label{e5}
\ee
in which
\begin{equation}
\begin{aligned}
\label{e6}
\textgoth{W} & =  \alpha (X+Y) \\
\textgoth{F} & =  - \frac{\Lambda}{3\alpha^2} +(Y^2-1)(1-2M\alpha Y + Q^2 \alpha^2 Y^2) \\
\textgoth{G} & =  (1 - X^2)(1 + 2M\alpha X + \alpha^2 Q^2 X^2).
\end{aligned}
\end{equation}
for the new parameters $\alpha, \Lambda, M$ and $Q$ corresponding to the acceleration, cosmological constant, mass and charge of the black hole. See also the discussion in Section \ref{conicity} about the physical meaning of the parameters.   
The transformation between the above coordinate systems can be derived as
\begin{equation}
\begin{aligned}
\label{e9}
y & =  \beta \gamma (Y+\delta)\\
x & =  \beta \gamma (X-\delta) \\
t & =   \frac{\gamma}{\beta} \tau \\
\Phi & =  \frac{\gamma}{\beta} \phi
\end{aligned}
\end{equation}
which is a generalization of the transformation introduced in \cite{Hong_2003}. The relation between the parameters of (\ref{e1}) and (\ref{e5}) is obtained from the determination of the triad $(\beta, \gamma, \delta)$ through
\be
\label{e13}
4\alpha^2Q^2 \delta^3 +6\alpha M \delta^2 + 2(1-\alpha^2 Q^2)\delta - 2\alpha M & = & 0\nonumber\\
\nonumber
\bigg( 3\alpha^2Q^2 \delta^4 +4\alpha M \delta^3 + (1-\alpha^2 Q^2)\delta^2 
\label{e14}
\left. +1 + \frac{\Lambda}{3\alpha^2}+\frac{1}{l^2\alpha^2}\right)^{-1/2}  & = & \beta \\
\label{e15}
(1-\alpha^2 Q^2 + 6\alpha M \delta + 6\alpha^2 Q^2 \delta^2 )^{1/2} & = & \gamma.\nonumber
\ee
There is at least one positive real root for $\delta$ in equation (\ref{e13}), which may be used to solve the subsequent equations for $\beta$ and $\gamma$. The remaining two roots are either imaginary (yielding non-physical spacetimes) or forbidden by the restriction $|al| \geq 1$ (see \cite{Krtou__2005} for more details). The determination of the triad allows us to find the constants in the system \eqref{e9} through
\be
\label{e16}
a & =&  \frac{\alpha}{\beta} \\
\label{e17}
q &  = & \frac{Q}{\gamma^2} \\
\label{e18}
m & = & \frac{M+2\alpha \delta Q^2}{\gamma^3}.
\ee
We note that the coordinate transformations must be such that the order of the roots of $G$ and $\textgoth{G}$ are preserved. 
We restrict our analysis to the interval $-1\le X\le 1$. 

Furthermore, we note that the rescaling of the cosmological constant in terms of $-l^2\Lambda = 3\tilde{c}$ induces different values of $\beta$ in equation (\ref{e14}) depending on the choice of $\tilde{c}$, and has impact on the relations between $(m,a)$ and $(M,\alpha )$. We can freely choose $\tilde{c}=1$, such that, equation (\ref{e14}) is the same as found in \cite{Hong_2003}. In turn, the asymptotic AdS behaviour can be recovered with the proper rescaling of the coordinates.\footnote{The shift in $a$ can be absorbed through the relation $\sin \chi = a^2l^2+\tilde{c}-1$, see (\ref{e26b}), preserving the conformal structure.}

We shall use the above metric forms in order to determine the conformal properties of the spacetime and draw Penrose diagrams.
However, to study the quasinormal modes through the field propagation in the boosted geometry, we will use the spherically symmetric coordinate system, in which the non-minimally coupled scalar field equation can be properly decoupled and integrated with the usual boundary conditions. 

The coordinate transformations from (\ref{e5}) to spherical coordinates were given in \cite{griffiths_podolsky_2009} as
\begin{equation}
\begin{aligned}
\label{e19}
X & =   - \cos \theta \\
Y &  =  \frac{1}{\alpha r} \\
\tau & =  \alpha t,
\end{aligned}
\end{equation}
leading the metric to the form
\be
ds^2 = \Omega^{-2} \left( - fdt^2 + \frac{1}{f}dr^2 + \frac{r^2}{P}d\theta^2 + r^2 P\sin^2 \theta d\phi^2 \right),
\label{e22}
\ee
with 
\be
\label{e23}
\Omega & =& \Omega (r,\theta):= 1- \alpha r \cos \theta \\
\label{24}
f & = & f(r) := \left(\frac{r^2-2Mr+Q^2}{r^2}\right)(1-\alpha^2 r^2 ) -\frac{\Lambda r^2}{3} \\
\label{e25}
P & = & P(\theta ) := 1 - 2\alpha M \cos \theta  + \alpha^2 Q^2 \cos^2 \theta.
\ee
which will be used in sections \ref{s4} and \ref{s5}.
We shall now revise some of the relevant geometric aspects the boosted Reissner-Nordtr\"om AdS spacetime. 
\section{Geometric properties of the spacetimes}\label{s3}

The BRNAdS spacetime is static, axially symmetric and has two Killing horizons defined by $f=0$ for small accelerations (i.e. for $a^2 -\frac{1}{l^2}<0$), corresponding to the Cauchy and event horizons. For large accelerations, a third null hypersurface is present corresponding to the acceleration horizon. The spacetime with large acceleration has a similar conformal structure as the case with $\Lambda \geq 0$ \cite{Dias_2003b} and will not be treated here. For studies of quasinormal modes of a scalar field in boosted RN black holes with an acceleration horizon, see \cite{Destounis2020pjk,destounis2020stability}.

\subsection{Conformal structure}

Penrose diagrams were already given in \cite{Krtou__2005, Dias_2003} with emphasis on cases with large accelerations. Here, we focus on cases with small accelerations and provide further details.

In order to study the conformal structure of the BRNAdS spacetime we will use the line element (\ref{e1}). In fact the radial coordinate in (\ref{e22}) is not complete in defining conformal infinity. Furthermore, in the regime of high $Q$ the interpretation of $r$ as the usual radial coordinate loses its meaning 
(together with the metric parameters in that coordinate system) and an analytical continuation to negative regions is needed (see also a discussion in \cite{griffiths_podolsky_2009}). 

In the metric (\ref{e1}), the spacetime admits the two null directions
\be
\label{e26}
n_1 = \partial_t - F\partial_y, \hspace{1.0cm} n_2 = \partial_t + F\partial_y
\ee
and the metric is of Petrov type D with the two Killing vectors $\partial_t$ and $\partial_\phi$.

In order to construct the conformal diagram for $a^2 < \frac{1}{l^2}$, which accommodates a single black hole with asymptotic AdS boundary, 
we first define an alternative coordinate 
\be 
z =  y\tan \chi,
\ee 
with the angle $\chi$ given by \cite{Krtou__2005}
\be
\label{e26b}
\sin \chi = al
\ee
and flat-conformal coordinates,
\begin{equation}
\begin{aligned}
\label{e27}
Z  &=  \int \frac{1}{\mathcal{F}}dz\\
T  &=  t \cot \chi,
\end{aligned}
\end{equation}
where
$$\mathcal{F}:= F \tan \chi = q^2l^{-2} \cos^2 \chi (z-z_c)(z-z_h)(z^2 + Az+B)$$
for a black hole with horizons at $z_c$ and $z_h$. In such case, $Z$ is expressed as
\be
\nonumber
Z = k_1 \arctan \left( \frac{A+2z}{\sqrt{4B-A^2}}\right) + k_2 \ln (z^2+Az+B) + 
\label{e29}
+k_c \ln (z_c-z)  + k_h \ln (z_h-z),
\ee
with
\begin{equation}
\begin{aligned}
\label{e30}
k_1 & =  \frac{(A^2-2B + 2z_c z_h + A(z_c+z_h))q^{-2}l^2\sec^2 \chi }{\sqrt{4B-A^2}(B+z_c(A+z_c))(B+z_h(A+z_h))} \\
k_2 & =   \frac{(A+z_c+z_h)q^{-2}l^2\sec^2 \chi }{2(B+z_c(A+z_c))(B+z_h(A+z_h))} \\
k_c & =   \frac{l^2\sec^2 \chi}{q^2(B+z_c(A+z_c))(z_c-z_h)} \\
k_h & =   \frac{l^2\sec^2 \chi}{q^2(B+z_h(A+z_h))(z_h-z_c)}.
\end{aligned}
\end{equation}
Finally, setting
\begin{equation}
\begin{aligned}
\label{e34}
u & = & Z+T \\
v & = & Z-T,
\end{aligned}
\end{equation}
we can prescribe the double-null compact coordinates $U$ and $V$,
\begin{equation}
\begin{aligned}
\label{e36}
U & = & 2\arctan \left[ (-1)^m \exp  \left( \frac{u}{\Delta} \right) \right] \\
V & = & 2\arctan \left[ (-1)^n \exp  \left( \frac{v}{\Delta} \right) \right],
\end{aligned}
\end{equation}
where $m$ and $n$ are integers and $\Delta$ is such that $\Delta = 2 |k_{c}|$ in the vicinity of $z_c$ and $z_{singularity}$ or  $\Delta = 2 |k_{h}|$ in the vicinity of $z_h$ and $z_\infty$. 
\begin{figure*}
\subfigure{\includegraphics[scale=0.9]{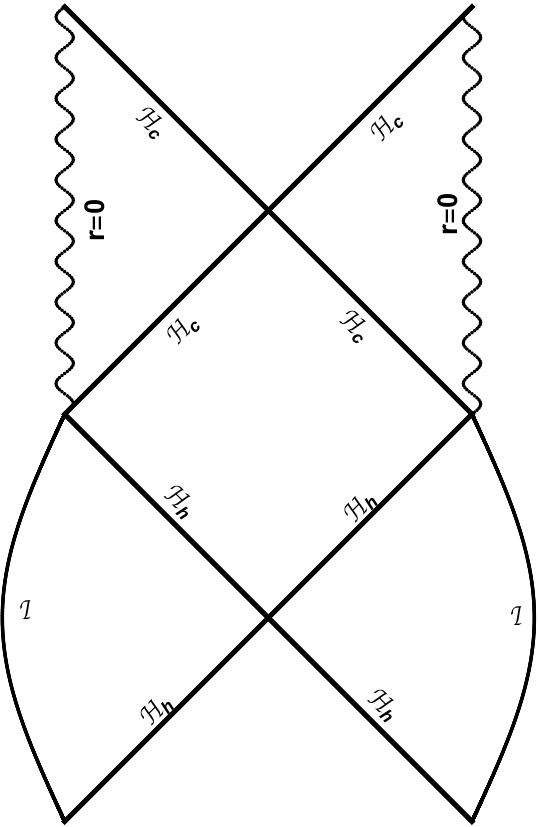}}\hskip 8ex
\subfigure{\includegraphics[scale=0.9]{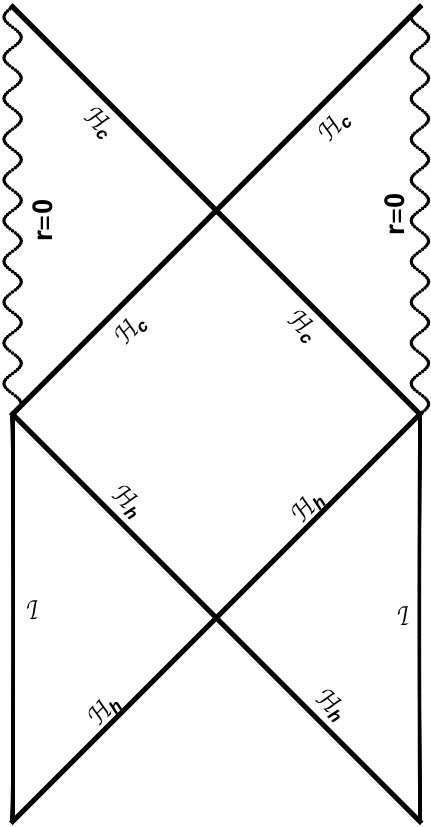}}\hskip 8ex
\subfigure{\includegraphics[scale=0.9]{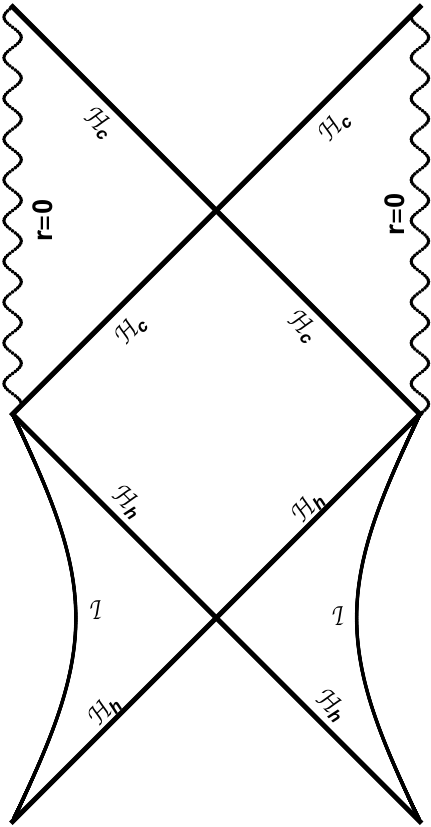}}\hskip 8ex
\caption{Conformal structure of the BRNAdS black hole. The shape of the asymptotic region depends on $\theta$ and the black hole parameters. We parametrize the middle panel with a straight line at infinity
and $\theta=\pi /2$. In the left and right panels $\theta = 0$ and $\theta = \pi$, respectively. ${\cal H}_h$ and ${\cal H}_c$ denote the event and Cauchy horizons, respectively.
}
\label{f1}
\end{figure*}
In turn the curvature invariants of metric (\ref{e1}) read 
\begin{equation}
\begin{aligned}
\label{e38}
\mathcal{R} & =  - \frac{12}{l^2}\\
\mathcal{R}_{\mu \nu}\mathcal{R}^{\mu \nu} & =  \frac{36+4a^8l^4q^4(x+y)^8}{l^4} \\
\mathcal{R}_{\mu \nu \sigma \alpha}\mathcal{R}^{\mu \nu \sigma \alpha} & =  \frac{24}{l^4} + 8a^6(x+y)^6(12mq^2(x-y)  +6m^2  + a^2q^4(7x^2-10xy+7y^2)),
\end{aligned}
\end{equation}
displaying a singularity at $y \rightarrow \infty$ or $r=0$ (see also \cite{Dias_2003}). At conformal infinity the spacetime has the same curvature invariants as the AdS spacetime with ${\cal R}$ being proportional to $\Lambda$. 

The above considerations allow us to draw the Penrose diagrams of Figure \ref{f1}. 

\subsection{Conicity, mass and charge}
\label{conicity}

We recall that the metric coefficient $g_{\theta \theta}$ must be positive. 
Such condition narrows down the coordinate range to the interval between the two real roots of $P$ which happens in the vicinity of $P=1$. Through these coordinates, we may see that the ratio circumference/radius along the axes $\theta =0$ and $\theta = \pi$ is not 2$\pi$ but a function of $P$. Then the excess or lack of angle $\phi$ is given by 
\be
\delta_{\theta_0} = 2\pi \left(1 - \lim_{\theta \rightarrow \theta_0} \frac{\sqrt{\frac{g_{\phi \phi}}{g_{\theta \theta}}}}{\theta-\theta_0}\right) =  2\pi (1 - CP(\theta_0)),
\label{e41}
\ee
where $C$ is a constant. By choosing e.g. $CP(\pi )=1$, the axis $\theta = \pi$ has no conical singularity, but $\theta = 0$ does (one can choose the other way around). This corresponds to a spacetime string or strut depending on which singularity is avoided \cite{Dias_2003}.

The constant $C$ is called conicity of the solution, being a fundamental constant of the spacetime that defines the range of $\phi$ within $[0, 2\pi C]$. Moreover, it relates to the presence of a thermodynamical quantity, the tension of the string \cite{Gregory_1995,Gregory_2017}
\be
\label{e42}
\mu \simeq \frac{M\alpha}{C} = \frac{M\alpha}{1+2M\alpha+Q^2\alpha^2}.
\ee
The string tension modifies non-trivially the first law of black hole thermodynamics as well as the entropy and, thus, the event horizon area formula (see e. g. \cite{Appels_2016, Appels_2017}).

The transformations of the previous section raise the question of the interpretation of the parameters $m, q$ and $M,Q$ and their relation to the mass and charge of the black holes. Recently, this issue has been addressed in \cite{Appels_2017} where a framework for the thermodynamics of a charged accelerating black hole has been put forward. To define the mass and charge of an accelerating black hole Gregory at al. \cite{Appels_2017} used the method of Ashtekar et al. of conformal completion \cite{ashtekar1981, ashtekar2000}. This involves defining the mass as an integral (over a sphere at conformal infinity) of the electric part of the Weyl tensor projected along the timelike conformal Killing vector. Their calculations give
\be
\label{e42a}
{\mathfrak M}=\frac{M}{C},~~~~~~{\mathfrak Q}=\frac{Q}{C},
\ee
which correspond to the mass and charge observed at infinity.

\section{Evolution equation and numerics}\label{s4}

In this section we consider the evolution equation for a non-minimally coupled scalar field on a BRNAdS background.
We show how the scalar field wave equation can be decoupled with a convenient transformation, leading to a system of two differential equations and to a double eigenvalue problem for the corresponding linear differential operators. The angular equation will be further analyzed in the boosted Schwarzschild AdS case where it reduces to a Heun differential equation \cite{ronveaux1995heun}.

\subsection{The eigenvalue problems}

We consider a massless neutral scalar field non-minimally coupled to the Ricci curvature scalar on a BRNAdS spacetime $({\cal M}, g)$, expressed through the matter field action as
\be
\label{e43}
\mathcal{S}_m = \int_{\cal M} d^4x \sqrt{-g}\left(\partial_\mu \Phi \partial^\mu \Phi - \xi\mathcal{R}\Phi^2 \right)
\ee
yielding the Klein-Gordon equation
\be
\label{e44}
\Box_g \Phi - \xi \mathcal{R}\Phi = 0,
\ee
where $\Box_g= g_{\alpha\beta} \nabla^\alpha\nabla^\beta$ is the wave operator with respect to the metric (\ref{e22}) whose components are denoted by $g_{\alpha\beta}$. We note that rigorous results about the existence and stability of solutions to the above equation have been proved in \cite{holzegel} for Schwarzschild AdS backgrounds and in \cite{kehle} for Reissner-Nordstr\"om AdS backgrounds.

We choose $6\xi=1$ which allows the separation of the wave equation into angular and radial-temporal parts\footnote{We note that this procedure would not be possible for the wave equation $\Box_g \phi=0$ which is not conformally invariant in this setting. See also \cite{papa2009} where the a conformally coupled scalar field is considered.}.
Expanding (\ref{e44}) and multiplying by the auxiliary function $r^2\Omega^{-2}$ we get
\begin{equation}
\begin{aligned}
&-\frac{\mathcal{R}r^2}{6\Omega^2}\Phi -\frac{r^2}{f}\partial_{tt}\Phi + \partial_r(r^2f)\partial_r \Phi + r^2f\partial_{rr} \Phi + \frac{1}{P\sin^2 \theta}\partial_{\phi \phi}\Phi \\
&+ \frac{\partial_\theta (P\sin \theta)}{\sin \theta}\partial_\theta \Phi + P \partial_{\theta \theta} \Phi - 2r^2f\Omega_r \partial_r \Phi - 2P\Omega_\theta \partial_\theta \Phi = 0
\label{e45}
\end{aligned}
\end{equation}
where 
$$\Omega_r = \frac{\partial_r \Omega}{\Omega}~~~{\text {and}}~~~\Omega_\theta =\frac{\partial_\theta \Omega}{\Omega}$$
and we recall that $\Omega$ is given by \eqref{e23}.
We now replace the first term using the identity
\be
\nonumber
\frac{\mathcal{R}r^2}{6\Omega^2}\Phi = 
\nonumber
 \left(\frac{\partial_\theta (P \sin \theta) \Omega_\theta}{\sin \theta} - 2P \Omega_\theta^2 + \frac{P \partial_{\theta \theta} \Omega }{\Omega}+ \frac{{\cal R}_{\tilde g}r^2}{6}
 +\partial_r (r^2f)\Omega_r - 2fr^2 \Omega_r^2+ \frac{fr^2 \partial_{rr} \Omega }{\Omega}\right) \Phi 
\label{e46}
\ee
in which 
\be
\nonumber
{\cal R}_{\tilde g}= -\frac{2f + 4r\partial_r f+r^2\partial_{rr} f}{r^2}- \frac{ 3\cot \theta \partial_\theta P + \partial_{\theta \theta} P - 2P}{r^2}
\label{e47}
\ee
represents the Ricci scalar of the metric $\tilde g_{\alpha\beta}$ conformal to $g_{\alpha
\beta}$ as given by (\ref{e22}), 
$$\tilde g_{\alpha\beta}=\Omega^2 g_{\alpha
\beta}.$$
Then equation (\ref{e45}) can be written as
\begin{align}
\label{e48}
& -\frac{r^2}{f}\partial_{tt}\Phi + \partial_r(r^2f)\partial_r \Phi + r^2f\partial_{rr} \Phi + 
 \frac{1}{P\sin^2 \theta}\partial_{\phi \phi}\Phi 
+ \frac{\partial_\theta (P\sin \theta)}{\sin \theta}\partial_\theta \Phi + P \partial_{\theta \theta} \Phi \nonumber\\
&- 2r^2f\Omega_r \partial_r \Phi- 2P\Omega_\theta \partial_\theta \Phi 
+\left(2P \Omega_\theta^2- \frac{\partial_\theta (P \sin \theta) \Omega_\theta}{\sin \theta} - \frac{P \partial_{\theta \theta} \Omega }{\Omega} - \frac{R_{\tilde g}r^2}{6}\right.\\
\nonumber
& \left.-\partial_r (r^2f)\Omega_r + 2fr^2 \Omega_r^2- \frac{fr^2 \partial_{rr} \Omega }{\Omega}\right)\Phi =0.
\end{align}
Now, multiplying (\ref{e48}) by $\Omega^{-1}$ and applying the transformation
\be
\label{e49}
\Phi = \Omega \Psi,
\ee
we get the separable form
\be
\nonumber
-\frac{r^2}{f} \partial_{tt} \Psi+ \partial_r \big( r^2f\partial_r \Psi \big) + \frac{1}{P \sin^2 \theta}\partial_{\phi \phi} \Psi +\frac{1}{\sin \theta}\partial_\theta (P \sin \theta \partial_\theta \Psi)\\
\nonumber
+\frac{1}{6}(2f + 4r\partial_r f+r^2\partial_{rr} f) \Psi+ \frac{1}{6}(3\cot \theta \partial_\theta P + \partial_{\theta \theta} P - 2P)\Psi = 0.
\label{e50}
\ee
Finally, taking the ansatz
\be
\label{e51}
\Psi = \frac{\psi (r,t) \zeta (\theta) e^{im\phi}}{r},
\ee
and tortoise-like coordinates for $r$ and $\theta$, 
\be
\label{e52}
dr_* =  \frac{1}{f}dr,~~~~~
d\Theta  =  \frac{1}{P\sin \theta} d\theta,
\ee
we obtain the {\em radial-temporal equation} and the {\em angular equation} as
\be
\label{e54}
\left( \frac{\partial^2}{\partial t^2}-\frac{\partial^2}{\partial r_*^2} + V \right)\psi=0 \\
\label{e55}
\left( \frac{\partial^2}{\partial \Theta^2}-m^2 + \vartheta \right)\zeta =0
\ee
with
\be
\label{e56}
V(r) & = & f\left(\frac{\lambda}{r^2} - \frac{f}{3r^2} + \frac{\partial_r f}{3r} - \frac{\partial_{rr} f}{6} \right) \\
\label{e57}
\vartheta(\theta) & = & P\sin^2\theta\left( \lambda - \frac{P}{3}+\frac{\cot \theta \partial_\theta P}{2} +\frac{\partial_{\theta \theta}  P}{6} \right)
\ee
where $\lambda$ and $m$ denote the eigenvalues associated to the operators in \eqref{e54} and \eqref{e55}, respectively. 
Recalling that the coordinate $\phi$ is periodic within $[0,2\pi C]$, then the eigenvalue $m$ is written as
$$m=m_0 C,$$ 
with $m_0 \in \mathbb{Z}$. The eigenvalue $\lambda$ of the angular part is expected to be near $\ell (\ell + 1)+1/3$, $\ell \in \mathbb{N}$, which represents the exact solution in spherical harmonics for $\alpha =0$.

The boundary conditions for the quasinormal problem are given by the usual plane wave solution near the event horizon and the constant solution at spatial infinity, since the potential tends to constant as $r\rightarrow \infty$, and we can write them as
\begin{equation}
\begin{aligned}
\label{e58}
\psi |_{r_*\rightarrow -\infty}& \rightarrow e^{-i\omega r_*}\\
\psi |_{r_*\rightarrow 0} &\rightarrow 0,
\end{aligned}
\end{equation}
although a secondary wavefront might be taken in limit $r_*\rightarrow 0$ \cite{Chen_2020}, since $V$ remains bounded there. In both cases though, the integration scheme we use assures the same set of quasinormal modes\footnote{It was noticed in \cite{Chen_2020} the presence of a secondary group of frequencies when the extra wavefront exists. That corresponds to $\psi\big|_{r_*\rightarrow 0} \sim const$ which brings no modification to the characteristic integration.  Also, the frequencies produced were in general very high for the imaginary part of the modes and far from the values obtained for $const=0$.}.

For the angular part, we consider boundary conditions compatible with the angular equation \eqref{e55}, maintaining $\zeta$ finite along the boundary values of the coordinate $\theta$,
\begin{equation}
\begin{aligned}
\label{e60}
\zeta | _{\Theta \rightarrow -\infty} &\rightarrow e^{m\Theta }\\
\zeta |_{\Theta \rightarrow \infty} &\rightarrow e^{-m\Theta}.
\end{aligned}
\end{equation}
We will employ different numerical methods to obtain the eigenvalues in each case.

For the radial-temporal component \eqref{e54}, we apply the characteristic integration in double null coordinates $\mathfrak{u}=t-r_*$ and $\mathfrak{v}=t+r_*$. We propagate a Gaussian package 
\be
\label{e62}
\psi\big|_{\mathfrak{v}=\mathfrak{v}_0} = e^{-\kappa(\mathfrak{u}-\mathfrak{u}_0)^2}  
\ee
along the $\mathfrak{u}$-$\mathfrak{v}$ diagram and collect the wave signal to be analyzed through the prony method \cite{Konoplya_2011}. The quasinormal frequencies are obtained altogether within this wave signal by applying such spectroscopic technique with an {\it a priori} number of overtones to be collected.

\subsection{The angular equation}

The angular part of the wave equation, differently from the radial-temporal part, has real eigenvalues which will be computed through a Frobenius expansion around one of the coordinate poles.

\subsubsection{Case $Q\ne 0$}

The Frobenius method starts by redefining the angular variable as
\be
\label{e63}
w=\frac{1-\cos \theta}{2},
\ee
in order to turn (\ref{e55}) into a more suitable form,
\be
\label{e64}
T(T\zeta')' - \frac{2}{3\alpha^2 Q^2}(6m^2+\alpha^2 Q^2 T(6\lambda + T''))\zeta=0,\hspace{0.4cm}
\ee
where the prime denotes a derivative with respect to $w$ and 
$$T(w) = 16\prod_{i=1}^4(w-w_i)$$
is a polynomial with the four regular singular points (in the sense of the theory of ordinary differential equations),
\be
\label{e65}
w_1=0, \hspace{0.5cm} w_2=1, \hspace{0.5cm} w_{3,4} = \frac{\alpha Q^2 - r_{3,4}}{2\alpha Q^2},
\ee
with $r_{3,4}= \frac{M \pm \sqrt{M^2-Q^2}}{2}$. Equation (\ref{e64}) has a fifth regular singularity $w_5$ at infinity. Such number of singularities does not allow to transform \eqref{e64} into a Heun differential equation with the usual field variables modifications \cite{Maier_2006, forsyth1900theory, Fiziev_2009, Horta_su_2018}. In such case, we are able to solve the problem only numerically with the method detailed in Appendix B which is adapted from \cite{Destounis2020pjk}.

\subsubsection{Case $Q= 0$}

A different situation occurs when $Q=0$ in which case the angular equation becomes
\be
T(T\zeta')' +4\Big(-m^2+  T\Big(\lambda +2\alpha M (1-2w) -\frac{1}{3}\Big)\Big)\zeta=0.
\label{e66}
\ee
Using now
\be
\nonumber
T(w)=-16\alpha M \prod_{i=1}^3(w-w_i),
\ee
with
\be
w_1=0, \hspace{0.8cm} w_2=1, \hspace{0.8cm} w_3= \frac{-1+2\alpha M}{4\alpha M},
\label{e68}
\ee
gives
\begin{equation}
\begin{aligned}
2bw(w-1)(w-w_3)\partial_w \Big( 2bw(w-1)(w-w_3)\partial_w \zeta \Big)
= \Big( m^2 + 4bw(w-1)(w-w_3)(c-bw)\Big)\zeta
\label{e67}
\end{aligned}
\end{equation}
with
\be
\nonumber
b=4\alpha M, \hspace{1.2cm} c=\lambda - \frac{1}{3} + 2\alpha M.
\ee
In this case the angular equation has four regular singular points and can be transformed into a Heun differential equation \cite{ronveaux1995heun}. Then, the solution to (\ref{e67}) can be written in terms of Heun general functions as 
\begin{equation}
\begin{aligned}
\zeta (w) = &C_1 \mathfrak{b}_+ HeunG \left( \frac{b-2}{2b} , \frac{c}{b} ,1,1, \frac{bm_0+b+2m_0-2}{b-2},m_0+1,w\right)\\
\label{e69}
+& C_2 \mathfrak{b}_- HeunG \left( \frac{b-2}{2b} , b_1,b_2,b_2, b_2,m_0+1,w \right)
\end{aligned}
\end{equation}
for
\begin{equation}
\begin{aligned}
\mathfrak{b}_\pm & =  w^{\pm \frac{m_0(b+2)}{2b-4}}(w-1)^{m_0/2}(2-b+2bw)^{\frac{bm_0}{2-b}} \\
b_1 & =  \frac{p_1m_0^2+p_2m_0 +2c(b-2)^2}{2b(b-2)^2} \\
p_1 & =  3b^3+10b^2+4b-8,\\
p_2 &=-3b^3+2b^2+12b-8 \\
\label{e70}
b_2 & =  \frac{(1-m_0)b-2m_0-2}{b-2}
\end{aligned}
\end{equation}
and $C_1$ and $C_2$ are real constants. The asymptotic behaviour of (\ref{e69}) requires $C_2=0$, as the companion function of $C_1$ is the only bounded solution satisfying conditions (\ref{e60}). Still, a Taylor series expansion of (\ref{e69}) around $w=0$ will be convergent whenever \cite{ronveaux1995heun}
\be
\label{e71}
\Bigg| \frac{b-2}{2b} \Bigg| < 1.
\ee
Since $b>0$, we obtain the convergence condition
\be
\label{e72}
\alpha M < \frac{1}{6}.
\ee
We note that this is the same convergence condition as obtained for the expansion of the wave function around a regular singular point in the case $Q=\Lambda=0$ of \cite{Destounis2020pjk}. A Frobenius method for the numerical computation of the eigenvalue $\lambda$ similar to the one developed in \cite{Destounis2020pjk} is detailed in Appendix \ref{apa}. In the limit of very small charges (e. g. $Q\lesssim 10^{-12}M$), this method yields a similar eigenvalue to the one found in \cite{Destounis2020pjk}.

\section{Quasinormal modes}\label{s5}

One might expect that the propagation of initial data with compact support for the scalar field in the black hole geometry will generate a decaying profile that decomposes into a tower of damped oscillations of quasinormal modes. It is well known that this is the case whenever $V>0$ \cite{Kokkotas:1999bd}. Such status would indicate the stability of the background with respect to the scalar perturbations. However, in our case there are regions where $V<0$. We will now investigate this issue by studying the Schwarzschild and Reissner-Nordström AdS accelerated black holes separately. 

\subsection{Boosted Schwarzschild AdS case}

In this case the potential (\ref{e56}) is positive for $r>r_h$ when $\lambda >1/3$. This occurs for every pair of eigenvalues $(\ell , m_0 )$ of the angular equation with $\ell >0$. In the special case $\ell = 0$ we have $\lambda < 1/3$ in general resulting in a potential partially negative for the region $r>r_h$ which may bring instabilities to the field evolution.  
However, as can be seen in Table \ref{tang1}, even for the highest values of the event horizon radius and acceleration, the deviation of $\lambda$ from $1/3$ is very mild. Since the change of sign in $V$ happens at $r = r_+ := \frac{6M}{1-3\lambda}$, this means that $V(r_*)<0$ on a small interval $(-\epsilon, 0)$, for $|\epsilon| \ll h$, where $h$ defines the grid step of integration. This fact turns unlikely the presence of instabilities in the field evolution. In fact, we did extensive numerical simulations of the scalar field evolution for small black holes and we have found, after the initial burst, the usual oscillatory phase that endures until late times. This indicates the spacetime stability with respect to the linear scalar field perturbation. Table \ref{t1} contains a representative sample of the quasinormal frequencies $\omega$ found.

Interestingly, as we increase the acceleration parameter, the change in $\Re (\omega )$ and $\Im (\omega)$ take opposite directions: the former increases while the latter decreases. This is also true if we consider increasing $|m_0|$ for fixed $\alpha$, $r_h$ and $\ell$. To illustrate this we can observe, in Table \ref{t1}, that every $\Im (\omega)_{m_0=0}$ is larger than $\Im (\omega)_{m_0=1}$ and that, on the contrary, $\Re (\omega)_{m_0=0}<\Re (\omega)_{m_0=\pm 1}$. 

\begin{figure*}
\subfigure{\includegraphics[scale=0.4]{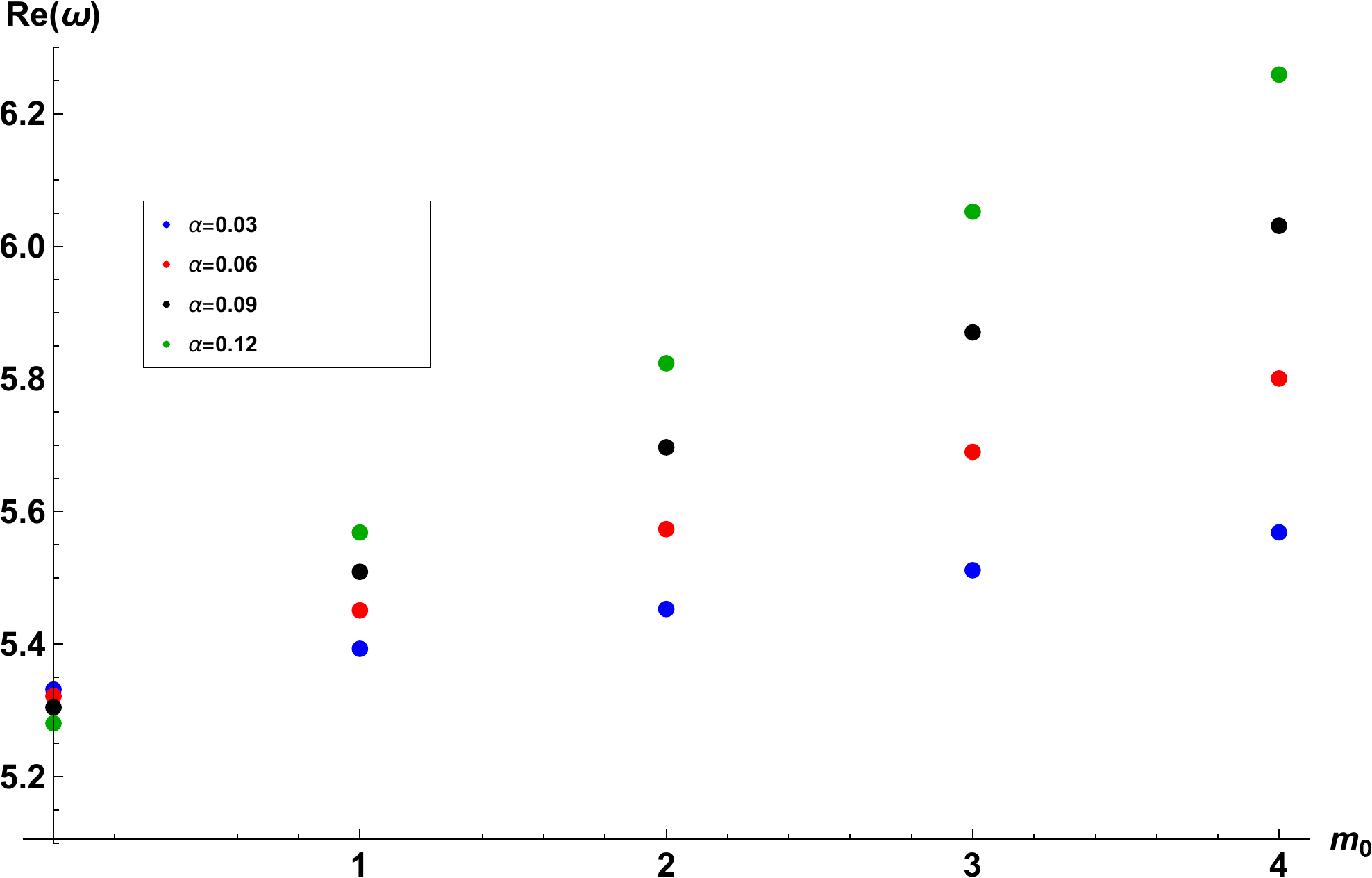}}\hskip 2ex
\subfigure{\includegraphics[scale=0.4]{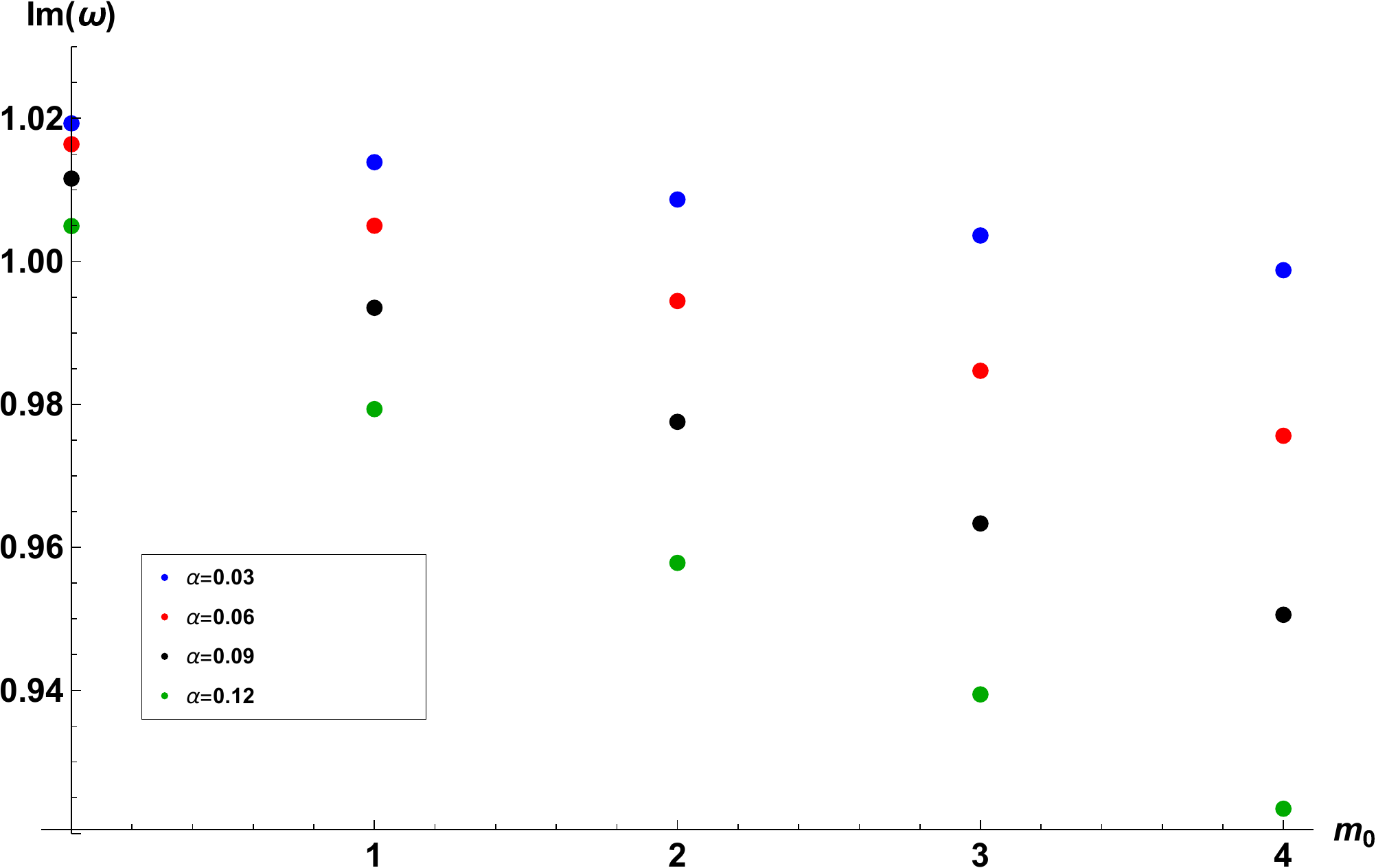}}\hskip 2ex
\caption{Quasinormal modes of the boosted Schwarzschild AdS black hole for $r_h=1$ and different $m_0$. Each eigenvalue $\lambda$ is given in Table \ref{ae2}  of Appendix B and corresponds to the fifth angular eigenvalue ($\ell = 4$) of each case.}
\label{fm2}
\end{figure*}

We have also computed the frequencies for the fifth eigenvalue, $\ell =4$, for different $m_0$ in the case $3r_h = -\Lambda = 3$ and for different acceleration parameters. The results are given in Figure \ref{fm2} confirming that, as we vary $m_0$, the increment of the field azimuthal angular momentum increases $\Re (\omega )$ and decreases $\Im (\omega )$, thus improving the quality factor of the spacetime
$$\mathbf{Q}=\frac{\Re (\omega )}{-\Im (\omega )}.$$ 
The variation in $\mathbf{Q}$ follows a similar pattern if we increase $\ell$. By comparing the last two columns with the first two columns of Table \ref{t1}, we observe that the change in $\mathbf{Q}$ for the second eigenvalue of angular momentum is of about $1\%$ for $m_0=0$ and about $10\%$ for $m_0=1$.

\begin{table*}
  \centering
 \caption{The first ($\ell =0$) angular eigenvalue $\lambda$ of the boosted Schwarzschild AdS black holes with $\Lambda=-3$.}
    \scalebox{0.75}{\begin{tabular}{cccccccccc}
    \hline
    $r_h$ & \multicolumn{2}{c}{$\alpha=0.03$} &   \multicolumn{2}{c}{$\alpha=0.06$} & \multicolumn{2}{c}{$\alpha=0.09$}	&   \multicolumn{2}{c}{$\alpha=0.12$}	\\
	\hline \hline
0.2	&	\multicolumn{2}{c}{0.333326843646}	&	\multicolumn{2}{c}{0.333307373544}	&	\multicolumn{2}{c}{0.333274919903}	&	\multicolumn{2}{c}{0.333229477512}	\\
0.4	&	\multicolumn{2}{c}{0.333301035946}	&	\multicolumn{2}{c}{0.333204107919}	&	\multicolumn{2}{c}{0.333042441470}	&	\multicolumn{2}{c}{0.332815856349}	\\
0.6	&	\multicolumn{2}{c}{0.333233421490}	&	\multicolumn{2}{c}{0.332933283827}	&	\multicolumn{2}{c}{0.332431707754}	&	\multicolumn{2}{c}{0.331726651464}	\\
0.8	&	\multicolumn{2}{c}{0.333074906497}	&	\multicolumn{2}{c}{0.332296910238}	&	\multicolumn{2}{c}{0.330991099266}	&	\multicolumn{2}{c}{0.329143397162}	\\
1.0	&	\multicolumn{2}{c}{0.332732202536}	&	\multicolumn{2}{c}{0.330915104778}	&	\multicolumn{2}{c}{0.327839878228}	&	\multicolumn{2}{c}{0.323432092330}	\\
    \hline  
    \end{tabular}}
  \label{tang1}
\end{table*}

\begin{table*}
  \centering
 \caption{Quasinormal modes of boosted Schwarzschild AdS black holes with $\Lambda=-3$.}
   \scalebox{0.73}{ \begin{tabular}{cccccccccc}
    \hline
     &  & \multicolumn{2}{c}{$\alpha=0.03$} &   \multicolumn{2}{c}{$\alpha=0.06$} & \multicolumn{2}{c}{$\alpha=0.09$}	&   \multicolumn{2}{c}{$\alpha=0.12$}	\\
$r_h$ & $( \ell ,m_0)$ & {$\Re(\omega)$} & {$-\Im(\omega)$} & {$\Re(\omega)$} & {$-\Im(\omega)$} & {$\Re(\omega)$} & {$-\Im(\omega)$} &  {$\Re(\omega)$} & {$-\Im(\omega)$}  \\ 
	\hline \hline
\vspace{0.3cm} 
0.2	& $(0,0)$	&	1.721852	&	0.176419	&	1.720124	&	0.175806	&	1.717236	&	0.174785	&	1.713179	&	0.173357	\\
&	$(1,0)$		&	2.762191	&	0.025072	&	2.759331	&	0.024902	&	2.754548	&	0.024623	&	2.747818	&	0.024245	\\
\vspace{0.3cm} 
&	$(1,\pm 1)$	&	2.769023	&	0.024648	&	2.773004	&	0.024061	&	2.775059	&	0.023375	&	2.775151	&	0.022597	\\
\vspace{0.3cm}
0.4	& $(0,0)$	&	1.631413	&	0.523901	&	1.630082	&	0.522541	&	1.627857	&	0.520273	&	1.624729	&	0.517097	\\
&	$(1,0)$		&	2.503694	&	0.313820	&	2.501580	&	0.312606	&	2.498048	&	0.310585	&	2.493085	&	0.307757	\\
\vspace{0.3cm}	
&	$(1,\pm 1)$	&	2.517637	&	0.311202	&	2.529609	&	0.307365	&	2.540282	&	0.302720	&	2.549619	&	0.297274	\\
\vspace{0.3cm}	
0.6	&	$(0,0)$		&	1.665875	&	0.863752	&	1.664954	&	0.861889	&	1.663413	&	0.858781	&	1.661244	&	0.854426	\\
&	$(1,0)$		&	2.437852	&	0.681911	&	2.436027	&	0.680042	&	2.432977	&	0.676926	&	2.428690	&	0.672563	\\
\vspace{0.3cm}	 
&	$(1,\pm 1)$	&	2.460361	&	0.677410	&	2.481504	&	0.670979	&	2.501855	&	0.663239	&	2.521374	&	0.654185	\\
\vspace{0.3cm}
0.8	&	$(0,0)$		&	1.762268	&	1.191167	&	1.761986	&	1.189024	&	1.761512	&	1.185451	&	1.760839	&	1.180448	\\
&	$(1,0)$		&	2.467364	&	1.036596	&	2.465721	&	1.034409	&	2.462967	&	1.030767	&	2.459081	&	1.025674	\\
\vspace{0.3cm}	
&	$(1,\pm 1)$	&	2.501639	&	1.029955	&	2.535487	&	1.020942	&	2.569435	&	1.010265	&	2.603456	&	0.997900	\\
\vspace{0.3cm}
1.0	& 	$(0,0)$		&	1.896277	&	1.511802	&	1.896969	&	1.509651	&	1.898121	&	1.506072	&	1.899724	&	1.501076	\\
&	$(1,0)$		&	2.546882	&	1.378140	&	2.545149	&	1.376025	&	2.542207	&	1.372522	&	2.537978	&	1.367663	\\
&	$(1,\pm 1)$	&	2.596948	&	1.368790	&	2.648150	&	1.356866	&	2.701079	&	1.343022	&	2.755756	&	1.327205	\\
\hline
    \end{tabular}}
  \label{t1}
\end{table*}

The presence of different quasinormal frequencies as we change $m_0$ indicates the existence of a fine structure for the scalar field, which is not present in the non-accelerated Schwarzschild AdS black hole (neither in the non-accelerated Reissner-Norstr\"om AdS geometry) and is a direct consequence of the presence of $\alpha$. Indeed, in the non-accelerated Schwarzschild black hole, each 'state' $\ell$ has $2\ell + 1$ degenerate substates with identical  spectrum, while in the boosted Schwarzschild  AdS, the fine structure changes such distribution as the change in $\lambda$, for each $m_0$ (for fixed $\ell$), is now very pronounced. This variation can be observed in the tables of Appendix \ref{apb} for the angular eigenvalues that we have computed.  

\subsection{Boosted Reissner-Nordström AdS case}

\begin{table*}
  \centering
 \caption{Quasinormal modes of the BRNAdS black holes with $\Lambda=-3$ and $\mathcal{Q}=0.1$.}
\scalebox{0.75}{
    \begin{tabular}{cccccccccc}
    \hline
     &  & \multicolumn{2}{c}{$\alpha=0.03$} &   \multicolumn{2}{c}{$\alpha=0.06$} & \multicolumn{2}{c}{$\alpha=0.09$}	&   \multicolumn{2}{c}{$\alpha=0.12$}	\\
$r_h$ & $( \ell ,m_0)$ & {$\Re(\omega)$} & {$-\Im(\omega)$} & {$\Re(\omega)$} & {$-\Im(\omega)$} & {$\Re(\omega)$} & {$-\Im(\omega)$} &  {$\Re(\omega)$} & {$-\Im(\omega)$}  \\ 
	\hline \hline
\vspace{0.3cm} 
0.2	& $(0,0)$	&	1.718881	&	0.177636	&	1.717162	&	0.177020	&	1.714290	&	0.175994	&	1.710254	&	0.174559	\\
	& $(1,0)$	&	2.759956	&	0.025368	&	2.757105	&	0.025197	&	2.752339	&	0.024917	&	2.745631	&	0.024535	\\
\vspace{0.3cm}
	& $(1,\pm 1)$	&	2.766871	&	0.024936	&	2.770947	&	0.024341	&	2.773105	&	0.023647	&	2.773309	&	0.022861	\\
\vspace{0.3cm}
0.4	& $(0,0)$	&	1.626331	&	0.527058	&	1.625017	&	0.525697	&	1.622819	&	0.523427	&	1.619729	&	0.520247	\\
	& $(1,0)$	&	2.499482	&	0.316147	&	2.497383	&	0.314931	&	2.493875	&	0.312904	&	2.488946	&	0.310068	\\
\vspace{0.3cm}
	& $(1,\pm 1)$	&	2.513610	&	0.313490	&	2.525787	&	0.309610	&	2.536680	&	0.304918	&	2.546253	&	0.299424	\\
\vspace{0.3cm}
0.6	& $(0,0)$	&	1.658217	&	0.868747	&	1.657321	&	0.866891	&	1.655822	&	0.863795	&	1.653711	&	0.859458	\\
	& $(1,0)$	&	2.432055	&	0.686018	&	2.430248	&	0.684153	&	2.427228	&	0.681044	&	2.422982	&	0.676691	\\
\vspace{0.3cm}
	& $(1,\pm 1)$	&	2.454942	&	0.681440	&	2.476503	&	0.674934	&	2.497304	&	0.667117	&	2.517307	&	0.657986	\\
\vspace{0.3cm}
0.8	& $(0,0)$	&	1.751439	&	1.198093	&	1.751188	&	1.195978	&	1.750766	&	1.192451	&	1.750165	&	1.187512	\\
	& $(1,0)$	&	2.459416	&	1.042430	&	2.457783	&	1.040266	&	2.455047	&	1.036663	&	2.451187	&	1.031625	\\
\vspace{0.3cm}
	& $(1.\pm 1)$	&	2.494402	&	1.035648	&	2.529034	&	1.026510	&	2.563833	&	1.015713	&	2.598778	&	1.003232	\\
\vspace{0.3cm}
1.0	& $(0,0)$	&	1.881849	&	1.520744	&	1.882570	&	1.518658	&	1.883767	&	1.515187	&	1.885435	&	1.510343	\\
	& $(1,0)$	&	2.536224	&	1.385794	&	2.534463	&	1.383741	&	2.531472	&	1.380343	&	2.527171	&	1.375635	\\
\vspace{0.3cm}
	& $(1,\pm 1)$	&	2.587522	&	1.376203	&	2.640081	&	1.364077	&	2.694505	&	1.350048	&	2.750822	&	1.334060	\\
\hline
    \end{tabular}}
  \label{t2}
\end{table*}

\begin{table*}
  \centering
 \caption{Quasinormal modes of the BRNAdS black holes with $\Lambda=-3$ and $\mathcal{Q}=0.7$.}
\scalebox{0.73}{
    \begin{tabular}{cccccccccc}
    \hline
     &  & \multicolumn{2}{c}{$\alpha=0.03$} &   \multicolumn{2}{c}{$\alpha=0.06$} & \multicolumn{2}{c}{$\alpha=0.09$}	&   \multicolumn{2}{c}{$\alpha=0.12$}	\\
$r_h$ & $( \ell ,m_0)$ & {$\Re(\omega)$} & {$-\Im(\omega)$} & {$\Re(\omega)$} & {$-\Im(\omega)$} & {$\Re(\omega)$} & {$-\Im(\omega)$} &  {$\Re(\omega)$} & {$-\Im(\omega)$}  \\ 
	\hline \hline
\vspace{0.3cm} 
0.2	& $(0,0)$	&	1.570213	&	0.273557	&	1.568928	&	0.272705	&	1.566780	&	0.271286	&	1.563760	&	0.269299	\\
	& $(1,0)$	&	2.636061	&	0.0499589	&	2.633737	&	0.0495998	&	2.629851	&	0.0490039	&	2.624387	&	0.0481749	\\
\vspace{0.3cm} 
	& $(1,\pm 1)$	&	2.646877	&	0.0489293	&	2.655441	&	0.0475567	&	2.662492	&	0.0459730	&	2.667984	&	0.0441919	\\
\vspace{0.3cm} 
0.4	& $(0,0)$	&	1.403189	&	0.786864	&	1.402657	&	0.785257	&	1.401766	&	0.782575	&	1.400506	&	0.778814	\\
	& $(1,0)$	&	2.305143	&	0.482371	&	2.303651	&	0.480928	&	2.301156	&	0.478521	&	2.297648	&	0.475151	\\
\vspace{0.3cm} 
	& $(1,\pm 1)$	&	2.328241	&	0.477328	&	2.350345	&	0.470802	&	2.371912	&	0.463277	&	2.392892	&	0.454756	\\
\vspace{0.3cm} 
0.6	& $(0,0)$	&	0	&	1.199676	&	0	&	1.198777	&	0	&	1.197278	&	0	&	1.195175	\\
	& $(1,0)$	&	2.190183	&	0.976503	&	2.189016	&	0.974854	&	2.187062	&	0.972107	&	2.184305	&	0.968263	\\
\vspace{0.3cm} 
	& $(1,\pm 1)$	&	2.232217	&	0.965776	&	2.274869	&	0.953254	&	2.318472	&	0.939462	&	2.362967	&	0.924370	\\
\vspace{0.3cm} 
0.8	& $(0,0)$	&	0	&	1.047781	&	0	&	1.046328	&	0	&	1.043885	&	0	&	1.040417	\\
	& $(1,0)$	&	2.154314	&	1.478695	&	2.152591	&	1.478218	&	2.149655	&	1.477470	&	2.145403	&	1.476530	\\
\vspace{0.3cm} 
	& $(1,\pm 1)$	&	2.225074	&	1.456170	&	2.299557	&	1.432795	&	2.378204	&	1.408642	&	2.460882	&	1.383604	\\
\vspace{0.3cm} 
1.0	& $(0,0)$	&	0	&	1.035007	&	0	&	1.032701	&	0	&	1.028663	&	***	&	***	\\
	& $(1,0)$	&	0	&	1.678920	&	0	&	1.669835	&	0	&	1.653840	&	***	&	***	\\
\vspace{0.3cm} 
	& $(1,\pm 1)$	&	0	&	1.769704	&	2.428626	&	1.900650	&	2.525291	&	1.870401	&	***	&	***	\\
\hline
    \end{tabular}}
  \label{t3}
\end{table*}

The BRNAdS black hole spacetime contains an event horizon and a Cauchy horizon. Here we exclude the range $\alpha^2 - \frac{\Lambda}{3} > 0$ for which an acceleration horizon forms. The extremal value of the charge defined for fixed $\alpha , \Lambda$ and $r_h$ is given by
\be 
\nonumber
Q_{max}=r_h\sqrt{1-r_h^2\Lambda\left(\frac{1-\alpha^2r_h^2/3}{(1-\alpha^2 r_h^2)^2}\right)}
\label{e73}
\ee
which in the special case $\alpha = 0$ matches the known expression of \cite{Wang_2004}. 

For convenience, the computation of the quasinormal frequencies is performed using the normalized charge
\be
\label{e74}
\mathcal{Q} = \frac{Q}{Q_{max}}
\ee
such that $\mathcal{Q}\in[0,1]$. The quasinormal modes for small and intermediate values of the charge are given in Table \ref{t2} and Table \ref{t3}. 
Note that for $\mathcal{Q}=0.1$ the respective frequencies are mildly different from those found in the non-charged case. In fact, the real and imaginary parts of the boosted Schwarzschild AdS spectra deviates slightly from the BRNAdS case with small charge (thus having a higher $\mathbf{Q}$). This feature is present for different pairs $(\ell , m_0)$.

In fact, the value of $\mathbf{Q}$ increases as we increase the black hole charge. In turn, the variation of the real and imaginary parts of the frequencies for small charges is again similar to that of a boosted Schwarzschild AdS black hole, while the variation of $\omega$ for increasing $\alpha$ is more pronounced for higher pairs $(\ell , m_0)$.  In the case of high charges, an interesting novelty is the presence of purely imaginary frequencies taking control of the field evolution, as they are smaller than the imaginary parts of the oscillatory modes. 

\begin{table*}
  \centering
 \caption{Subdominant oscillatory quasinormal modes of the BRNAdS black holes of Table \ref{t3}.}
    \scalebox{0.75}{\begin{tabular}{cccccccccc}
    \hline
     &  & \multicolumn{2}{c}{$\alpha=0.03$} &   \multicolumn{2}{c}{$\alpha=0.06$} & \multicolumn{2}{c}{$\alpha=0.09$}	&   \multicolumn{2}{c}{$\alpha=0.12$}	\\
$r_h$ & $( \ell ,m_0)$ & {$\Re(\omega)$} & {$-\Im(\omega)$} & {$\Re(\omega)$} & {$-\Im(\omega)$} & {$\Re(\omega)$} & {$-\Im(\omega)$} &  {$\Re(\omega)$} & {$-\Im(\omega)$}  \\ 
	\hline \hline
0.6	& $(0,0)$	&	1.390444	&	1.326985	&	1.390886	&	1.325262	&	1.391619	&	1.322387	&	1.392637	&	1.318355	\\
0.8	& $(0,0)$	&	1.461326	&	1.856400	&	1.463267	&	1.855531	&	1.466474	&	1.854304	&	1.471059	&	1.852342	\\
1.0	& $(0,0)$	&	1.567971	&	2.384519	&	1.572088	&	2.386000	&	1.578328	&	2.389188	&	***	&	***	\\
	& $(1,0)$	&	2.164401	&	1.999024	&	2.159934	&	2.002625	&	2.197220	&	1.989364	&	***	&	***	\\
	& $(1,\pm 1)$	&	2.273826	&	1.954646	&	***	&	***	&	***	&	***	&	***	&	***	\\
\hline
    \end{tabular}}
  \label{t4}
\end{table*}

The existence of purely imaginary frequencies may be related to a second family of quasinormal modes called near-extremal modes. In a non-accelerated Reissner-Nordstr\"om geometry such family is prominent, controlling the field evolution for 'very' near-extremal values of charge and being an important aspect in numerical studies of the strong cosmic censorship conjecture \cite{Cardoso_2018, Destounis_2019}. In the BRNAdS black hole, the presence of such damped purely imaginary decay is noticed for much smaller values of the black hole charge. This seems to indicate the presence of the near-extremal family, as long as the oscillatory family has an increasing imaginary part (with increasing $r_h$) that suprasedes the near-extremal modes. In this case, the secondary oscillatory mode can be acquired through the prony method by collecting 'higher overtones' in the field profile evolution, as it is shown in Table \ref{t5}. The frequencies of the subdominant oscillatory family, in the case of $\mathcal{Q}=0.7$, can be seen in Table \ref{t4}.

The quest of stability of the BRNAdS black hole is similar to the previous Schwarzschild AdS case: For $\ell >0$ the potential is positive outside the event horizon, thus generating a stable field profile decay, i.e. a composition of towers of quasinormal modes. Such behaviour was found numerically in our search with different geometry parameters which resulted in a field evolution constrained to that tower of quasi-eigenstates belonging to the different parameter families.

The special case $\ell =0$ results in $\lambda$-eigenvalues that allow for $V<0$ when $r>r_h$. The signal change in $V$ occurs at
\be
\label{e75}
r_\pm = \frac{3M \pm \sqrt{9M^2 + 6Q^2(1-\alpha^2Q^2 - 3\lambda)}}{1-\alpha^2Q^2 - 3\lambda}.
\ee
In general, far from the near-extremal regime, we will have $V (r>r_+)<0$ and $V(r_h<r<r_+)>0$. As in the boosted Schwarzschild AdS spacetime, the transition point $r_+$ is high enough to let $|r_*(r_+)| \ll h$, where $h$ defines the grid step of integration. This represents again a very small  $r_*$ interval, $(r_*(r_+),0)$, where the potential is negative and indicates the stability of the field evolution substantiated by our numerical data.
\begin{figure*}
\subfigure{\includegraphics[scale=0.4]{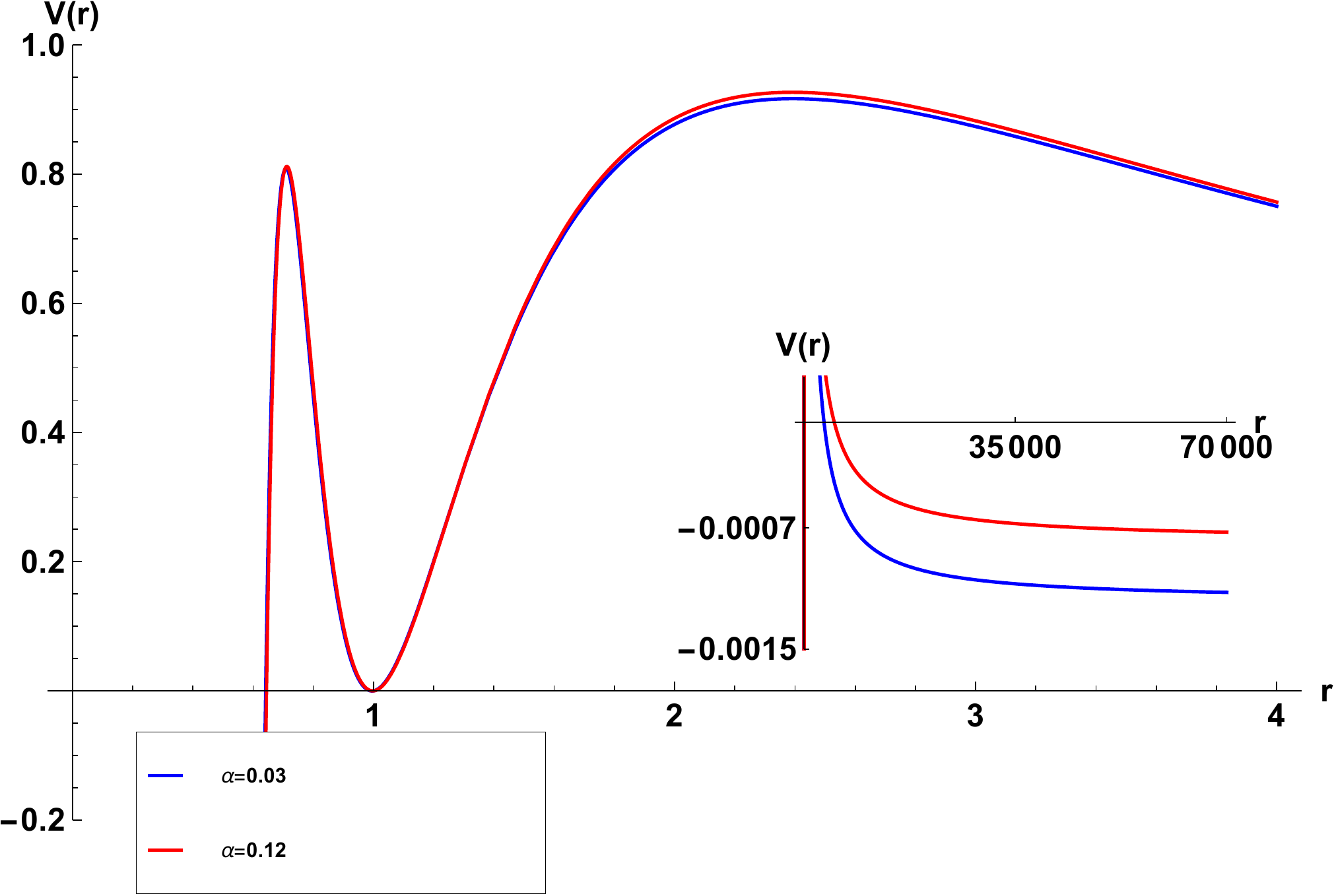}}\hskip 2ex
\subfigure{\includegraphics[scale=0.4]{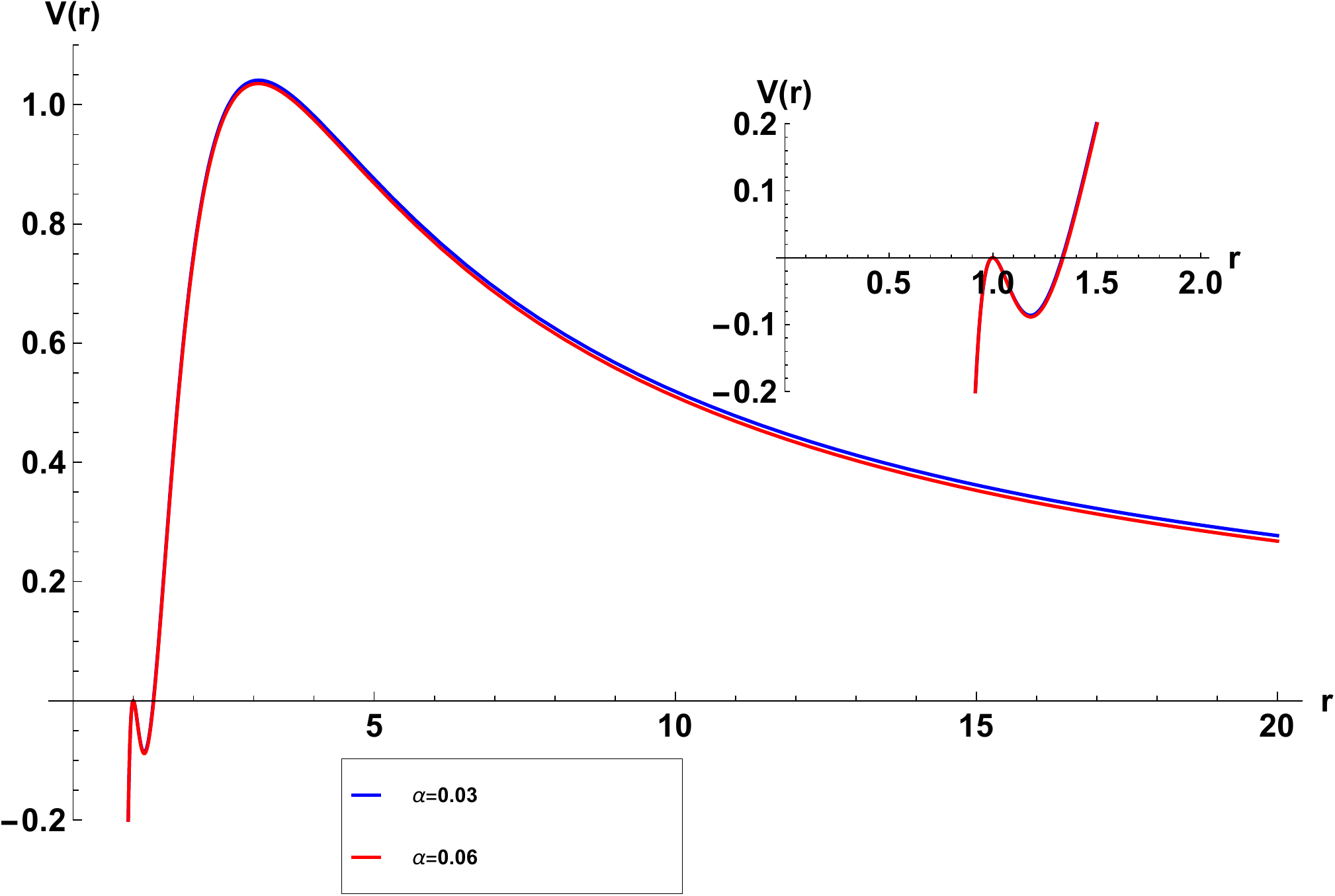}}\hskip 2ex
\caption{Scalar field potentials for the BRNAdS black hole with parameters $r_h=-\Lambda/3 =1$ and $\mathcal{Q}=0.7$ (left), $\mathcal{Q}=0.999$ (right) with $\ell = 0$. The regions where $V$ is negative is limited to: a small interval in $r_*$, for large $r$ as displayed in the left panel;  a small well in the near-extremal black hole as shown in the right panel; a small interval for large $r$ also in the near-extremal case. This situation does not yield unstable field evolutions in any case.}
\label{f2}
\end{figure*}
The behaviour of the frequencies relative to the variation of $m_0$ is similar to the boosted Schwarzschild AdS case, i.e. we find a fine structure for both real and imaginary parts of $\omega$, allowing the increment of $\mathbf{Q}$ for increasing $m_0$ (also observed for increasing $\ell$).

\subsubsection{Near-extremal modes}

We start by introducing the quantity
\be
\eta = 10^3(1-\mathcal{Q})
\ee
as a more suitable variable such that $\eta\in [1,10]$.

\begin{table}
  \centering
\caption{Near-extremal quasinormal modes with $3r_h=-\Lambda=3$.}
    \scalebox{0.75}{\begin{tabular}{cccccc}
    \hline
     &  & \multicolumn{2}{c}{$\alpha=0.03$} &   \multicolumn{2}{c}{$\alpha=0.06$} 	\\
$\eta$ & $( \ell ,m_0)$ & {$\Re(\omega)$} & {$-\Im(\omega)$} & {$\Re(\omega)$} & {$-\Im(\omega)$} \\ 
	\hline \hline
\vspace{0.3cm}
1	& $(0,0)$	&	0	&	0.003825	&	0	&	0.003747	\\
	& $(1,0)$	&	0	&	0.008092	&	0	&	0.007999	\\
\vspace{0.3cm}
	& $(1,\pm 1)$	&	0	&	0.008756	&	0	&	0.009408	\\
\vspace{0.3cm}
4	& $(0,0)$	&	0	&	0.016608	&	0	&	0.016390	\\
	& $(1,0)$	&	0	&	0.031966	&	0	&	0.031602	\\
\vspace{0.3cm}
	& $(1,\pm 1)$	&	0	&	0.034577	&	0	&	0.037142	\\
\vspace{0.3cm}
7	& $(0,0)$	&	0	&	0.030203	&	0	&	0.029872	\\
	& $(1,0)$	&	0	&	0.056053	&	0	&	0.055420	\\
\vspace{0.3cm}
	& $(1,\pm 1)$	&	0	&	0.060608	&	0	&	0.065089	\\
\vspace{0.3cm}
10	& $(0,0)$	&	0	&	0.044233	&	0	&	0.043800	\\
	& $(1,0)$	&	0	&	0.080236	&	0	&	0.079339	\\
	& $(1,\pm 1)$	&	0	&	0.086726	&	0	&	0.093115	\\
\hline
    \end{tabular}}
  \label{t5}
\end{table}
In the near-extremal regime, in addition to the region where $r\in (r_+,\infty )$ we also have a negative potential for $r\in (r_h, r_-)$, i.e. $V<0$ for $r_*\in (-\infty , r_*(r_-))$. Nevertheless, only a very small deep of the potential is observed that is about one order of magnitude smaller than its peak in the region $r\in[r_-,r_+]$. This then indicates the numerical stability of the solution. Figure \ref{f2} displays the cases with qualitatively different potentials.

In turn, Table \ref{t5} contains results of the computation of the near-extremal quasinormal modes. The field profile is dominated by purely imaginary decaying modes for both $\ell=0$ and $\ell = 1$, as shown for example for $\eta \leq 10$ and $\mathcal{Q}\geq 0.99$. 

The exact expression for the near-extremal family found in \cite{Destounis2020pjk} reproduces similar results as those displayed in Table \ref{t5} within a deviation of $5\%$. The agreement of the mode values in Table \ref{t5} with the results of \cite{Destounis2020pjk} indicates that they represent the same family of modes which are also present in Reissner-Nordstr\"om and Reissner-Nordstr\"om dS black holes.

\section{Final Remarks}\label{s6}

Accelerated black holes have been an active area of research with interesting recent developments leading to a better understanding of their physical and mathematical properties \cite{Appels_2016, frost2020lightlike, Destounis2020pjk, zhang2020shadows, lim2020null}.

In this paper, we have studied the numerical stability of accelerated AdS black holes against  linear scalar perturbations. In order to do that we have considered a non-minimally coupled scalar field propagating on AdS black holes with small accelerations and computed the field evolution and their quasinormal modes.

For the boosted Schwarzschild AdS black hole we have found, through characteristic numerical integration, that only oscillatory quasinormal modes are present in the perturbation spectra and that the scalar field perturbation decays in time, for the range of accelerations considered. Furthermore, the oscillatory frequencies display a fine structure relative to the azimuthal angular momentum $m_0$ of the field. For a specific eigenstate $\ell$, different $m_0$ render different modes $\omega$ with increasing $\Re (\omega)$ and decreasing $\Im (\omega)$ as $m_0$ increases. This is a new behaviour peculiar to boosted black holes. In fact, for the non-accelerated AdS black holes any eigenstate $\ell$ is degenerated in $m$ and produces the same quasinormal mode.

For the boosted Reissner-Nordtr\"om AdS black hole, a similar fine structure is exhibited with respect to $m_0$. In this case, however, the existence of a Cauchy horizon gives rise to field profiles controlled by purely imaginary solutions for high values of the charge. 
As a consequence of the small deeps in the scalar potentials, together with the small intervals for $r_*$ where $V(r)<0$, the evolution of the scalar perturbation is observed to be stable.

In conclusion, our results strongly suggest that these boosted AdS black holes with small acceleration are stable against small linear scalar perturbations.
\\

{\noindent \bf{Acknowledgements.}} The authors are grateful to Kyriakos Destounis for helpful discussions. FCM thanks support from FCT/Portugal through CAMGSD, IST-ID, project UIDB/04459/2020 and UIDP/04459/2020 as well as CMAT, Univ. Minho, through project UIDB/00013/2020 and UIDP/00013/2020 and FEDER Funds COMPETE.

\appendix
\section{Coefficients of the Frobenius method for the scalar angular equation with $Q=0$}\label{apa}

We apply the Frobenius method (detailed in \cite{Destounis2020pjk}) to the angular equation (\ref{e67}) and we describe briefly  how the respective expansion coefficients are obtained. In order to do that we consider the ansatz
\be
\label{ap3}
\nonumber
\zeta (w)= w^\delta (w-1)^\gamma \sum_{n=0}^{+\infty} A_n w^n,
\ee
with $\delta, \gamma\in \mathbb{C}$, 
into (\ref{e67}) to obtain the recurrence relation 
\begin{align}
\nonumber
A_n=&\Delta_n\sum_{i=0}^{n-1}\Big((v_{2+i}+\tau_{2+i})(i+1-n-\delta) - (t_{1+i}
+\omega_{1+i}+u_{1+i})\\
+& s_{3+i}(-\delta^2 + \delta (3+2i-2n) 
-(n-1-i)(n-2-i)\Big)A_{n-i-1},\nonumber
\end{align} 
with
\be
\nonumber
\Delta_n=\frac{1}{u_0 + \tau_1(\delta + n) + s_2 (\delta^2 + \delta (2n-1)+n^2-n)},
\ee
and the terms $v_n,\tau_n , t_n, \omega_n, u_n$ and $s_n$ are the expansion coefficients of the equation 
which, with the given ansatz, are written as
\be
\nonumber
\omega & := & 16 b^2 \gamma (\gamma - 1) w^2(w-w_3)^2 = \sum_{n=0}^{\nu} \omega_n w^n, \\
\nonumber
v & := & 32 b^2 \gamma w^2(w-1)(w-w_3)^2 = \sum_{n=0}^{\nu} v_n w^n, \\
\nonumber
s & := & T^2 = \sum_{n=0}^{\nu} s_n w^n, \\
\nonumber
t & := & -4T'b\gamma w(w-w_3) = \sum_{n=0}^{\nu} t_n w^n,\\
\nonumber
\tau & := & TT' = \sum_{n=0}^{\nu} \tau_n w^n, \\
\nonumber
u & := & -4 \left( m_0^2(1+b/2)^2 +T(c-bw) \right)  = \sum_{n=0}^{\nu} u_n w^n,
\ee
where $\nu \in \mathbb{N}$ and $b,c$ and $T$ are given in subsection 4.2.2.

The boundary conditions (\ref{e60}) can be rewritten in terms of the coordinate $w$ as
\be
\zeta\Big|_{w\rightarrow 0} & = & A_0w^\delta, \nonumber\\
\zeta\Big|_{w\rightarrow 1} & = & \sum_{n=0}^{\nu}A_n(w-1)^{\gamma},\nonumber
\ee
and determine the exponents of our ansatz as
\be
\delta & = & \frac{1+2\alpha M}{1-2\alpha M}\gamma\nonumber\\
\gamma & = & \frac{m_0}{2}.\nonumber
\ee
The method requires that $A_0$ and $\sum_{n=0}^\nu A_n$ are constant and remain bounded. Then, the eigenvalue $\lambda$ is the one for which the series converges faster. The Frobenius method then follows as in the Appendix B of \cite{Destounis2020pjk} using the coefficients derived above.
\section{Angular eigenvalues}\label{apb}

The angular eigenvalues are listed in tables \ref{ae1} and \ref{ae2} for the boosted Schwarzschild AdS case and in tables \ref{ae3}, \ref{ae4} and \ref{ae5}
for the BRNAdS case.

\begin{table*}
  \centering
 \caption{The second ($\ell =1$) angular eigenvalues $\lambda$ of the boosted Schwarzschild AdS black hole with $\Lambda=-3$.}
\scalebox{0.75}{    \begin{tabular}{ccccccccccc}
    \hline
    $r_h$ & $(\ell , m_0)$ & \multicolumn{2}{c}{$\alpha=0.03$} &   \multicolumn{2}{c}{$\alpha=0.06$} & \multicolumn{2}{c}{$\alpha=0.09$}	&   \multicolumn{2}{c}{$\alpha=0.12$}	\\
	\hline \hline
0.2	& $(1 , 0)$ &	\multicolumn{2}{c}{2.333298289031}	&	\multicolumn{2}{c}{2.333193150587}	&	\multicolumn{2}{c}{2.333017901376}	&	\multicolumn{2}{c}{2.332772513678}	\\
\vspace{0.3cm} 
	& $(1 , \pm 1)$ &	\multicolumn{2}{c}{2.352100113297}	&	\multicolumn{2}{c}{2.370960670720}	&	\multicolumn{2}{c}{2.389915327738}	&	\multicolumn{2}{c}{2.408964400969}	\\
0.4	& $(1 , 0)$ &	\multicolumn{2}{c}{2.333158927615}	&	\multicolumn{2}{c}{2.332635518848}	&	\multicolumn{2}{c}{2.331762531208}	&	\multicolumn{2}{c}{2.330539001733}	\\
\vspace{0.3cm} 
	& $(1 , \pm 1)$ &	\multicolumn{2}{c}{2.375327006500}	&	\multicolumn{2}{c}{2.417792644473}	&	\multicolumn{2}{c}{2.460737190225}	&	\multicolumn{2}{c}{2.504167616545}	\\
0.6	& $(1 , 0)$ &	\multicolumn{2}{c}{2.332793811025}	&	\multicolumn{2}{c}{2.331173092362}	&	\multicolumn{2}{c}{2.328464689225}	&	\multicolumn{2}{c}{2.324657677389}	\\
\vspace{0.3cm} 
	& $(1 , \pm 1)$ &	\multicolumn{2}{c}{2.407500556252}	&	\multicolumn{2}{c}{2.483154962052}	&	\multicolumn{2}{c}{2.560346525295}	&	\multicolumn{2}{c}{2.639126560839}	\\
0.8	& $(1 , 0)$ &	\multicolumn{2}{c}{2.331937839414}	&	\multicolumn{2}{c}{2.327736825751}	&	\multicolumn{2}{c}{2.320686176298}	&	\multicolumn{2}{c}{2.310710590694}	\\
\vspace{0.3cm} 
	& $(1 , \pm 1)$ &	\multicolumn{2}{c}{2.453306519460}	&	\multicolumn{2}{c}{2.577205005199}	&	\multicolumn{2}{c}{2.705243211037}	&	\multicolumn{2}{c}{2.837644842922}	\\
1.0	& $(1 , 0)$ &	\multicolumn{2}{c}{2.330087286568}	&	\multicolumn{2}{c}{2.320275866010}	&	\multicolumn{2}{c}{2.303673700564}	&	\multicolumn{2}{c}{2.279920427719}	\\
	& $(1 , \pm 1)$ &	\multicolumn{2}{c}{2.517761087135}	&	\multicolumn{2}{c}{2.711482557976}	&	\multicolumn{2}{c}{2.915172999773}	&	\multicolumn{2}{c}{3.129533412902}	\\
    \hline  
    \end{tabular}}
  \label{ae1}
\end{table*}
\begin{table*}
  \centering
 \caption{Angular eigenvalues $\lambda$ with $\ell =4$ of the boosted Schwarzschild AdS black hole with $3r_h=\Lambda=-3$.}
\scalebox{0.75}{\begin{tabular}{cccccccccc}
    \hline
    $m_0$ & \multicolumn{2}{c}{$\alpha=0.03$} &   \multicolumn{2}{c}{$\alpha=0.06$} & \multicolumn{2}{c}{$\alpha=0.09$}	&   \multicolumn{2}{c}{$\alpha=0.12$}	\\
	\hline \hline
0	&	\multicolumn{2}{c}{20.30572099701}	&	\multicolumn{2}{c}{20.22226231855}	&	\multicolumn{2}{c}{20.08104555132}	&	\multicolumn{2}{c}{19.87872040093}	\\
1	&	\multicolumn{2}{c}{20.87611371905}	&	\multicolumn{2}{c}{21.42902918386}	&	\multicolumn{2}{c}{21.99951026422}	&	\multicolumn{2}{c}{22.59604586787}	\\
2	&	\multicolumn{2}{c}{21.43937370937}	&	\multicolumn{2}{c}{22.60338669832}	&	\multicolumn{2}{c}{23.83468102299}	&	\multicolumn{2}{c}{25.14323594138}	\\
3	&	\multicolumn{2}{c}{21.99553895303}	&	\multicolumn{2}{c}{23.74601976879}	&	\multicolumn{2}{c}{25.59052162022}	&	\multicolumn{2}{c}{27.53470285859}	\\
4	&	\multicolumn{2}{c}{22.54466781250}	&	\multicolumn{2}{c}{24.85790253689}	&	\multicolumn{2}{c}{27.27215921520}	&	\multicolumn{2}{c}{29.78713644778}	\\
    \hline  
    \end{tabular}}
  \label{ae2}
\end{table*}
\begin{table*}
  \centering
 \caption{Angular eigenvalues $\lambda$ of the BRNAdS black hole with $\Lambda=-3$ and $\mathcal{Q}=0.1$.}
\scalebox{0.75}{    \begin{tabular}{ccccccccccc}
    \hline
    $r_h$ & $(\ell , m_0)$ & \multicolumn{2}{c}{$\alpha=0.03$} &   \multicolumn{2}{c}{$\alpha=0.06$} & \multicolumn{2}{c}{$\alpha=0.09$}	&   \multicolumn{2}{c}{$\alpha=0.12$}	\\
	\hline \hline
0.2	& $(0 , 0)$ &	\multicolumn{2}{c}{0.333326837517}	&	\multicolumn{2}{c}{0.333307349027}	&	\multicolumn{2}{c}{0.333274864737}	&	\multicolumn{2}{c}{0.333229379426}	\\
	& $(1 , 0)$ &	\multicolumn{2}{c}{2.333298040894}	&	\multicolumn{2}{c}{2.333192158012}	&	\multicolumn{2}{c}{2.333015667982}	&	\multicolumn{2}{c}{2.332768542947}	\\
\vspace{0.3cm} 
	& $(1 , \pm 1)$ &	\multicolumn{2}{c}{2.352303887057}	&	\multicolumn{2}{c}{2.371372590208}	&	\multicolumn{2}{c}{2.390539801594}	&	\multicolumn{2}{c}{2.409805874504}	\\
0.4	& $(0 , 0)$ &	\multicolumn{2}{c}{0.333300916993}	&	\multicolumn{2}{c}{0.333203631726}	&	\multicolumn{2}{c}{0.333041368601}	&	\multicolumn{2}{c}{0.332813945442}	\\
	& $(1 , 0)$ &	\multicolumn{2}{c}{2.333157148535}	&	\multicolumn{2}{c}{2.332628399331}	&	\multicolumn{2}{c}{2.331746500271}	&	\multicolumn{2}{c}{2.330510472275}	\\
\vspace{0.3cm} 
	& $(1 , \pm 1)$ &	\multicolumn{2}{c}{2.375871987345}	&	\multicolumn{2}{c}{2.418907374042}	&	\multicolumn{2}{c}{2.462447057555}	&	\multicolumn{2}{c}{2.506498648789}	\\
0.6	& $(0 , 0)$ &	\multicolumn{2}{c}{0.333232588715}	&	\multicolumn{2}{c}{0.332929944209}	&	\multicolumn{2}{c}{0.332424161459}	&	\multicolumn{2}{c}{0.331713155034}	\\
	& $(1 , 0)$ &	\multicolumn{2}{c}{2.332785718749}	&	\multicolumn{2}{c}{2.331140664571}	&	\multicolumn{2}{c}{2.328391505356}	&	\multicolumn{2}{c}{2.324527017068}	\\
\vspace{0.3cm} 
	& $(1 , \pm 1)$ &	\multicolumn{2}{c}{2.408665900195}	&	\multicolumn{2}{c}{2.485572648034}	&	\multicolumn{2}{c}{2.564107786727}	&	\multicolumn{2}{c}{2.644327115676}	\\
0.8	& $(0 , 0)$ &	\multicolumn{2}{c}{0.333071229416}	&	\multicolumn{2}{c}{0.332282115738}	&	\multicolumn{2}{c}{0.330957483809}	&	\multicolumn{2}{c}{0.329082802532}	\\
	& $(1 , 0)$ &	\multicolumn{2}{c}{2.331909007110}	&	\multicolumn{2}{c}{2.327620960995}	&	\multicolumn{2}{c}{2.320423445562}	&	\multicolumn{2}{c}{2.310238346293}	\\
\vspace{0.3cm} 
	& $(1 , \pm 1)$ &	\multicolumn{2}{c}{2.455526720737}	&	\multicolumn{2}{c}{2.581892598674}	&	\multicolumn{2}{c}{2.712663759721}	&	\multicolumn{2}{c}{2.848083771123}	\\
1.0	& $(0 , 0)$ &	\multicolumn{2}{c}{0.332719915444}	&	\multicolumn{2}{c}{0.330865382867}	&	\multicolumn{2}{c}{0.327725787649}	&	\multicolumn{2}{c}{0.323223985944}	\\
	& $(1 , 0)$ &	\multicolumn{2}{c}{2.330001714700}	&	\multicolumn{2}{c}{2.319930221691}	&	\multicolumn{2}{c}{2.302883047186}	&	\multicolumn{2}{c}{2.278446512474}	\\
	& $(1 , \pm 1)$ &	\multicolumn{2}{c}{2.521647942458}	&	\multicolumn{2}{c}{2.719867695779}	&	\multicolumn{2}{c}{2.928727407306}	&	\multicolumn{2}{c}{3.148993761679}	\\
    \hline  
    \end{tabular}}
  \label{ae3}
\end{table*}
\begin{table*}
  \centering
 \caption{Angular eigenvalues $\lambda$ of the BRNAdS black hole with $\Lambda=-3$ and $\mathcal{Q}=0.7$.}
\scalebox{0.75}{    \begin{tabular}{ccccccccccc}
    \hline
    $r_h$ & $(\ell , m_0)$ & \multicolumn{2}{c}{$\alpha=0.03$} &   \multicolumn{2}{c}{$\alpha=0.06$} & \multicolumn{2}{c}{$\alpha=0.09$}	&   \multicolumn{2}{c}{$\alpha=0.12$}	\\
	\hline \hline
0.2	& $(0 , 0)$ &	\multicolumn{2}{c}{0.333324773122}	&	\multicolumn{2}{c}{0.333299091318}	&	\multicolumn{2}{c}{0.333256284397}	&	\multicolumn{2}{c}{0.333196346487}	\\
	& $(1 , 0)$ &	\multicolumn{2}{c}{2.333276571169}	&	\multicolumn{2}{c}{2.333106277538}	&	\multicolumn{2}{c}{2.332822431005}	&	\multicolumn{2}{c}{2.332424995816}	\\
\vspace{0.3cm} 
	& $(1 , \pm 1)$ &	\multicolumn{2}{c}{2.362097897588}	&	\multicolumn{2}{c}{2.391196700671}	&	\multicolumn{2}{c}{2.420632595437}	&	\multicolumn{2}{c}{2.450408417681}	\\
0.4	& $(0 , 0)$ &	\multicolumn{2}{c}{0.333282840912}	&	\multicolumn{2}{c}{0.333131285290}	&	\multicolumn{2}{c}{0.332878430787}	&	\multicolumn{2}{c}{0.332523882596}	\\
	& $(1 , 0)$ &	\multicolumn{2}{c}{2.333004974607}	&	\multicolumn{2}{c}{2.332019423693}	&	\multicolumn{2}{c}{2.330375252847}	&	\multicolumn{2}{c}{2.328070070650}	\\
\vspace{0.3cm} 
	& $(1 , \pm 1)$ &	\multicolumn{2}{c}{2.402121884881}	&	\multicolumn{2}{c}{2.472784748118}	&	\multicolumn{2}{c}{2.545369724045}	&	\multicolumn{2}{c}{2.619925174347}	\\
0.6	& $(0 , 0)$ &	\multicolumn{2}{c}{0.333137622108}	&	\multicolumn{2}{c}{0.332549129966}	&	\multicolumn{2}{c}{0.331563746663}	&	\multicolumn{2}{c}{0.330174503791}	\\
	& $(1 , 0)$ &	\multicolumn{2}{c}{2.332100326764}	&	\multicolumn{2}{c}{2.328393392034}	&	\multicolumn{2}{c}{2.322188588900}	&	\multicolumn{2}{c}{2.313445356211}	\\
\vspace{0.3cm} 
	& $(1 , \pm 1)$ &	\multicolumn{2}{c}{2.465010534629}	&	\multicolumn{2}{c}{2.603302572682}	&	\multicolumn{2}{c}{2.748536027677}	&	\multicolumn{2}{c}{2.901045550449}	\\
0.8	& $(0 , 0)$ &	\multicolumn{2}{c}{0.332701705859}	&	\multicolumn{2}{c}{0.330793750908}	&	\multicolumn{2}{c}{0.327569297922}	&	\multicolumn{2}{c}{0.322958059610}	\\
	& $(1 , 0)$ &	\multicolumn{2}{c}{2.329482746515}	&	\multicolumn{2}{c}{2.317857309942}	&	\multicolumn{2}{c}{2.298230820803}	&	\multicolumn{2}{c}{2.270208175180}	\\
\vspace{0.3cm} 
	& $(1 , \pm 1)$ &	\multicolumn{2}{c}{2.563543502614}	&	\multicolumn{2}{c}{2.812908360172}	&	\multicolumn{2}{c}{3.082822247896}	&	\multicolumn{2}{c}{3.374677803926}	\\
1.0	& $(0 , 0)$ &	\multicolumn{2}{c}{0.331561686530}	&	\multicolumn{2}{c}{0.326155642110}	&	\multicolumn{2}{c}{0.316615165824}	&	\multicolumn{2}{c}{***}	\\
	& $(1 , 0)$ &	\multicolumn{2}{c}{2.322824812111}	&	\multicolumn{2}{c}{2.290797459472}	&	\multicolumn{2}{c}{2.234454422975}	&	\multicolumn{2}{c}{***}	\\
	& $(1 , \pm 1)$ &	\multicolumn{2}{c}{2.712510606271}	&	\multicolumn{2}{c}{3.140238163558}	&	\multicolumn{2}{c}{3.628888161327}	&	\multicolumn{2}{c}{***}	\\
    \hline  
    \end{tabular}}
  \label{ae4}
\end{table*}
\begin{table*}
  \centering
 \caption{Angular eigenvalues $\lambda$ of the BRNAdS black hole with $3r_h=\Lambda=-3$.}
\scalebox{0.75}{    \begin{tabular}{cccccc}
    \hline
    $\eta$ & $(\ell , m_0)$ & \multicolumn{2}{c}{$\alpha=0.03$} &   \multicolumn{2}{c}{$\alpha=0.06$}		\\
	\hline \hline
1	& $(0 , 0)$ &	\multicolumn{2}{c}{0.329111161147}	&	\multicolumn{2}{c}{0.316010922864}	\\
	& $(1 , 0)$ &	\multicolumn{2}{c}{2.308616801645}	&	\multicolumn{2}{c}{2.232119643296}	\\
\vspace{0.3cm} 
	& $(1 , \pm 1)$ &	\multicolumn{2}{c}{2.923676195498}	&	\multicolumn{2}{c}{3.622431056504}	\\
4	& $(0 , 0)$ &	\multicolumn{2}{c}{0.329147462042}	&	\multicolumn{2}{c}{0.316163105766}	\\
	& $(1 , 0)$ &	\multicolumn{2}{c}{2.308824301404}	&	\multicolumn{2}{c}{2.232987160996}	\\
\vspace{0.3cm} 
	& $(1 , \pm 1)$ &	\multicolumn{2}{c}{2.921133606250}	&	\multicolumn{2}{c}{3.616534976067}	\\
7	& $(0 , 0)$ &	\multicolumn{2}{c}{0.329183475855}	&	\multicolumn{2}{c}{0.316314034822}	\\
	& $(1 , 0)$ &	\multicolumn{2}{c}{2.309030216868}	&	\multicolumn{2}{c}{2.23384772208}	\\
\vspace{0.3cm} 
	& $(1 , \pm 1)$ &	\multicolumn{2}{c}{2.918599959560}	&	\multicolumn{2}{c}{3.610661601828}	\\
10	& $(0 , 0)$ &	\multicolumn{2}{c}{0.329219204255}	&	\multicolumn{2}{c}{0.316463713017}	\\
	& $(1 , 0)$ &	\multicolumn{2}{c}{2.309234557044}	&	\multicolumn{2}{c}{2.234701378731}	\\
	& $(1 , \pm 1)$ &	\multicolumn{2}{c}{2.916075243898}	&	\multicolumn{2}{c}{3.604810894210}	\\
\hline  
    \end{tabular}}
  \label{ae5}
\end{table*}
\newpage
\bibliographystyle{JHEP}
\bibliography{references}
\end{document}